\documentclass[twocolumn]{aastex62}
\usepackage{amsmath,epsfig,url,natbib}
\usepackage{graphicx,epstopdf,float,color}
\usepackage{courier}
\usepackage[hang,flushmargin]{footmisc}
\usepackage{hyperref}
\usepackage[flushleft]{threeparttable}
\hypersetup{colorlinks=true, citecolor=blue, filecolor=magenta,
  urlcolor=cyan, linkcolor=blue}
\def\lsim{\mathrel{\rlap{\lower4pt\hbox{\hskip1pt$\sim$}}
    \raise1pt\hbox{$<$}}}                
\def\gsim{\mathrel{\rlap{\lower4pt\hbox{\hskip1pt$\sim$}}
    \raise1pt\hbox{$>$}}}                

\begin{document}
\title{\MakeUppercase{Massive molecular gas reservoir in a luminous
  submillimeter galaxy during cosmic noon}}
\author{Bin Liu}
\affiliation{Department of Astronomy, Beijing Normal University, Beijing 100875, China}
\affiliation{Department of Physics and Astronomy, University of
  California, Irvine, CA92697, USA}
\author{N. Chartab}
\affiliation{Department of Physics and Astronomy, University of
  California, Irvine, CA92697, USA}
\author{H. Nayyeri}
\affiliation{Department of Physics and Astronomy, University of
  California, Irvine, CA92697, USA}
\author{A. Cooray}
\affiliation{Department of Physics and Astronomy, University of
  California, Irvine, CA92697, USA}
\author{C. Yang}
\affiliation{Joint ALMA Observatory, Alonso de C$\acute{\rm o}$rdova 3107, Vitacura 763-0355, Santiago, Chile}
\affiliation{European Southern Observatory, Alonso de C$\acute{\rm o}$rdova 3107, Vitacura, Casilla 19001, Santiago de Chile, Chile}
\author{D.A Riechers}
\affiliation{Cornell University, Space Sciences Building, Ithaca, NY 14853, USA}
\author{M. Gurwell}
\affiliation{Center for Astrophysics $|$ Harvard $\&$ Smithsonian, 60 Garden Street, Cambridge, MA 02138, USA}
\author{Zong-hong Zhu}
\affiliation{Department of Astronomy, Beijing Normal University, Beijing 100875, China}
\affiliation{School of Physics and Technology, Wuhan University, Wuhan 430072, China}
\author{S. Serjeant}
\affiliation{School of Physical Sciences, The Open University, Milton Keynes, MK7 6AA, UK}
\author{E. Borsato}
\affiliation{Dipartimento di Fisica e Astronomia `G. Galilei', Università di Padova, Vicolo dell'Osservatorio 3, I-35122, Padova, Italy}
\author{M. Negrello}
\affiliation{School of Physics and Astronomy, Cardiff University, The Parade, Cardiff, CF24 3AA, UK}
\author{L. Marchetti}
\affiliation{Department of Astronomy, University of Cape Town, Private Bag X3, Ronderbosch, 7701 Cape Town, South Africa}
\affiliation{INAF - Istituto di Radioastronomia, via Gobetti 101, I-40129 Bologna, Italy}
\affiliation{School of Physical Sciences, The Open University, Milton Keynes, MK7 6AA, UK}
\author{E.M. Corsini}
\affiliation{Dipartimento di Fisica e Astronomia `G. Galilei', Università di Padova, Vicolo dell'Osservatorio 3, 35122, Padova, Italy}
\affiliation{ INAF - Osservatorio Astronomico di Padova, vicolo dell’Osservatorio 5, I-35122 Padova, Italy}
\author{P. van der Werf}
\affiliation{Leiden University, Leiden Observatory, PO Box 9513, NL-2300 RA, Leiden, The Netherlands}

\begin{abstract}

We present multiband observations of an extremely dusty star-forming lensed galaxy (HERS1) at $z=2.553$. High-resolution maps of HST/WFC3, SMA, and ALMA show a partial Einstein ring with a radius of $\sim$3$^{\prime\prime}$. The deeper HST observations also show the presence of a lensing arc feature associated with a second lens source, identified to be at the same redshift as the bright arc based on a detection of the [NII] 205 $\mu$m emission line with ALMA. A detailed model of the lensing system is constructed using the high-resolution HST/WFC3 image, which allows us to study the source-plane properties and connect rest-frame optical emission with properties of the galaxy as seen in submillimeter and millimeter wavelengths. Corrected for lensing magnification, the spectral energy distribution fitting results yield an intrinsic star formation rate of about $1000\pm260$\ $M_{\odot}$ yr$^{-1}$, a stellar mass ${ M_*}=4.3^{+2.2}_{-1.0}\times10^{11} \ {\rm M_{\odot}}$, and a dust temperature $T_{\rm d}=35^{+2}_{-1}$ K. The intrinsic CO emission line ($J_{\rm up}=3,4,5,6,7,9$) flux densities and CO spectral line energy distribution are derived based on the velocity-dependent magnification factors. We apply a radiative transfer model using the large velocity gradient method with two excitation components to study the gas properties. The low-excitation component has a gas density $n_{\rm H_2}=10^{3.8\pm0.6}$ cm$^{-3}$ and kinetic temperature $T_{\rm k}=18^{+7}_{-5}$ K, and the high-excitation component has $n_{\rm H_2}=10^{3.1\pm0.4}$ cm$^{-3}$ and $T_{\rm k}=480^{+260}_{-220}$ K. Additionally, HERS1 has a gas fraction of about $0.19\pm0.14$ and is expected to last 100 Myr. These properties offer a detailed view of a typical submillimeter galaxy during the peak epoch of star formation activity.
\end{abstract}



\section{Introduction}

The cold molecular gas (traced by millimeter (mm) observations of the molecular
CO) is the key fuel for active star formation in galaxies \citep{carilli2013}. The fraction of
the molecular gas reservoir that ends up in new stars (what is usually referred to as
the star formation efficiency) depends on parameters such as the
fragmentation and chemical compositions of the gas and is diminished by phenomena
that disperse the gas and prevent the collapse, such as feedback from an active nucleus (AGN) or
star-formation-driven winds \citep{Bigiel2008, Sturm2011, Swinbank2011, Fu2012}. Existing evidence suggests AGN activity is the dominant mechanism in quenching star formation at high redshifts, specifically in the most extreme
environments \citep{Cicone2014}. On the other hand, the UV emission from newly born hot
stars in star-forming regions ionizes the surrounding gas, generating a
wealth of recombination nebular emission lines. The presence and
intensity of these lines reveal valuable information on the physics of
ionized gas surrounding these regions \citep{Coil2015,Kriek2015, Shapley2015}. 

The most intense sites of star formation activities in the universe at high redshift happen in the
gas-rich dusty star-forming galaxies \citep{Casey2014}. These heavily dust-obscured systems are often discovered in the millimeter wavelengths and are
believed to be the progenitors of the most massive red galaxies found
at lower redshifts \citep{Toft2014}. Despite many efforts, the physics of the
ionized gas (chemical composition, spatial extent, and relative line abundance) is still
poorly understood for these star-forming factories of the
universe. Recent resolved studies of ionized gas at
high redshift mostly focus on normal star-forming galaxies and miss
this hidden population of starbursting systems \citep{Genzel2011, Forster2014}. 

Submillimeter galaxies (SMGs) are
more efficient in turning gas into stars than normal star-forming galaxies
at the same epoch (i.e., Lyman-break Galaxies (LBGs) and $BzK$-selected galaxies), similar to
local ultraluminous infrared galaxies (ULIRGs) \citep{Papadopoulos2012}. Recent studies suggest that the LBG selected star forming galaxies
might be fundamentally different from that of SMGs whereas the
former is usually characterized by ${\rm SFR<100}\,M_{\odot}\ {\rm yr}^{-1}$ the latter dominates the high-SFR
end, and these are believed to be driven by different star formation
mechanisms \citep{Casey2016}. Using unlensed SMGs,  \cite{Menendez2013} showed that the star
formation scale in these high-redshift systems seems to be very different from that
of local starburst and are more extended over 2-3 kpc scales.

\begin{figure}[htbp]
\centering
\includegraphics[trim=5.5cm 0cm 5.5cm 0cm, scale=1]{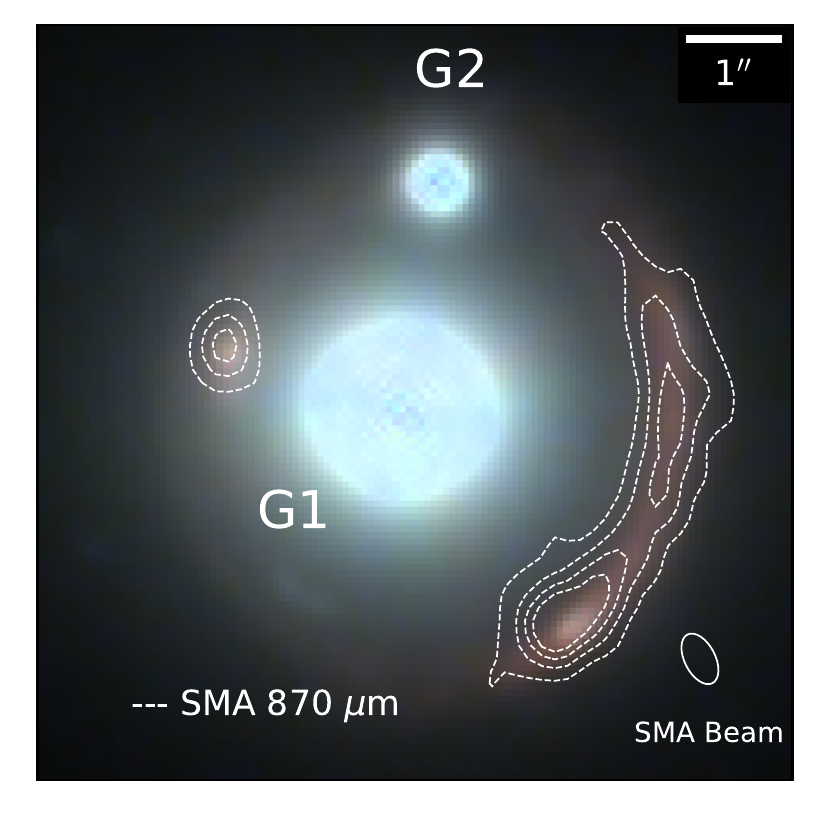}
\caption{Three-color image of HERS1 adopting HST/WFC3 F110W (blue), F125W (green), and F160W (red) with submillimeter array (SMA) $870\ \mu$m  contours overlaid. The SMA contours start from 9$\sigma$ and increase in steps of 9$\sigma$ with $\sigma=305 \ \mu$Jy\ beam$^{-1}$ . The beam size is shown in the bottom right. Two deflecting galaxies at $z=0.202$ \citep{Geach2015} are marked as G1 and G2.}
\label{fig:Fig three color}
\end{figure}

Through Herschel wide-area surveys, we have now identified hundreds of extremely bright submillimeter sources
($S_{500{\rm \mu m}} \geq 100$~mJy) at high redshifts. After removing nearby
contaminants, such bright 500\,$\mu$m sources are either gravitationally
lensed SMGs or multiple SMGs blended within the 18$^{\prime\prime}$ Herschel
PSF \citep{Negrello2007, Negrello2010,Negrello2017} with most turning out to be strongly lensed SMGs in our
high-resolution follow-up observations \citep{Fu2012,Bussmann2013,Wardlow2013,Timmons2015}. For this study we have
selected a very bright Keck/NIRC2-observed Einstein-ring-lensed
SMG at $z = 2.553$ (HERS J020941.1+001557 designated as HERS1 hereafter; Figure \ref{fig:Fig three color}). This target is identified from the Herschel Stripe
82 survey \citep{Viero2014} covering 81 deg$^2$ with  Herschel/SPIRE instrument at 250, 350, and 500\,$\mu$m.
 HERS1 at $z=2.553$ is the brightest galaxy in Herschel
  extra-galactic maps with
$S_{\rm 500\; \mu m} = 717 \pm 8$\,mJy. The galaxy was first identified  as a gravitationally lensed radio source by two foreground galaxies at $z=0.202$ in a citizen science project (\citealt{Geach2015}) to an
Einstein-ring with a radius $\sim$ 3$^{\prime\prime}$. HERS1 has also been selected by other wide-area surveys such as Planck and ACT and has extensive follow-up observations from CFHT and HST in
the near-infrared along with ancillary observations by JVLA, SCUBA-2, and ALMA with a CO redshift from the Redshift Search Receiver of Large millimeter telescope (LMT) and independently from H$\alpha$ using the IRCS on Subaru \citep{Geach2015, Harrington2016, su2017}. Here we report new data from Keck/NIRC2 Laser Guide Star Adaptive Optics (LGS-AO) imaging in $H$ and $K_s$ bands, HST/WFC3 F125W, Submillimeter Array (SMA), and ALMA.
The wealth of multiband data combined with high-resolution deep imaging provides a unique opportunity to study the physical properties of HERS1 as an extremely bright SMG during the peak epoch of star formation activity.

This paper is organized as follows: In Section \ref{section:data} we present observations and data reduction as well as previous archival data. In Section \ref{section:model}, we describe the lens-modeling procedures and reconstructed source-plane images of the high-resolution observations. We then show the source properties including the CO spectral line energy distribution (SLED), delensed CO spectral lines, large velocity gradient (LVG) modeling as well as infrared spectral energy distribution (SED) fitting in Section \ref{section:physical}. Finally, we summarize our results in Section \ref{section:summary}.

Throughout this paper, we assume a standard flat-$\Lambda$CDM cosmological model with $H_0=70$ km\ s$^{-1}$ \ Mpc$^{-1}$ and $\Omega_{\Lambda}=0.7$. All magnitudes are in the AB system.

\section{Data}
\label{section:data}

\subsection{Hubble Space Telescope WFC3 imaging}
HERS1 was observed on 2018 September 02 under GO program 15475 in Cycle 25 with two orbits (PI: Nayyeri). We used the WFC3 F125W filter with a total exposure time of 5524 s. The data were reduced by the HST pipeline, resulting in a scale of $0^{\prime \prime}.128$ pixel$^{-1}$. The photometry was performed following the WFC3 handbook \citep{Rajan2011}. 

\subsection{Keck Near-IR imaging}
The near-IR data of HERS1 was observed with the KECK/NIRC2 Adaptive Optics system on 2017 August 27 (PID: U146; PI: Cooray). The $H$ and $K_s$-band filters were used at 1.60 $\mu m$ and 2.15 $\mu$m respectively.
The observations were done with a custom nine-point dithering pattern for sky subtraction with 120 s ($H$ band) and 80 s ($K_s$ band) exposures per frame. Each frame has a scale of $0^{\prime \prime}.04$ pixel$^{-1}$ adopting the wide camera in imaging mode. The data were reduced by a custom {\sc IDL} routine \citep{Fu2012}.

\begin{table*}[hbt!]
    \centering
        \caption{Summary of Emission-line observations}
        \label{tab:observation}
        \centering
    \begin{tabular}{ccccccc}
    \hline
    \hline
    Science Goal & $\nu_{\rm obs}$ & t$_{\rm on}$ & Beam Size & Beam & rms   & References \\ [1ex]
                 & (GHz)           & (min) & ($\prime \prime$)   & ($^{\circ}$)  & (mJy \ beam$^{-1}$ )(bandwidth)   &  \\[1ex]
    \hline
         
    CO(3-2)      & 97.3         & 62    & 0.45$\times$0.32 &  89       & 0.250 (11.8 MHz)       &  - \\[1ex]
    CO(4-3)      & 129.7        & 108   & 0.27$\times$0.21 &  88       & 0.150 (23.2 MHz)       &  1  \\[1ex]
    CO(5-4)      & 162.2        & 10    & 10.50$\times$5.90&  -80      & 1.800 (221.6 MHz)      &  This work\\[1ex]
    CO(6-5)      & 194.6        & 18    & 8.14$\times$5.20 &  76       & 1.400 (266.1 MHz)      &  This work\\[1ex]
    CO(7-6)      & 227.0        & 12    & 6.88$\times$4.64 &  87       & 1.040 (310.6 MHz)      &  This work\\[1ex]
    CO(9-8)      & 291.8        & 15    & 5.28$\times$3.32 &  76       & 1.200 (399.1 MHz)      &  2\\[1ex]
    CI(1-0)      & 138.5        & 108   & 0.27$\times$0.21 &  88       & 0.150 (23.2 MHz)       &  1 \\[1ex]
    CI(2-1)      & 227.8        & 12    & 6.88$\times$4.64 &  87       & 1.040 (310.6 MHz)      &  This work \\[1ex]
    [NII] 205 $\mu$m     & 411.2        & 42    & 0.32$\times$0.25 &  80       & 3.000 (15.6 MHz)       &  3 \\[1ex]
    \hline

    \end{tabular}
    \tablerefs{(1): \cite{Geach2018}; (2): \cite{Riechers2021}; (3): \cite{Doherty2020}.}
\end{table*}

\subsection{Submillimeter array}
 
Observations of HERS1 were obtained in three separate configurations
of the SMA as described below.  In each
configuration, observations of HERS1 were obtained in tracks shared
with a second target. Generally, seven of the eight SMA antennas participated in the observations, except in one case (see below).  The SMA operates in
double-sideband mode, with sideband separation handled through a
standard phase-switching procedure within the correlator \citep{Ho2004}. During the period, the new SMA SWARM correlator was
expanding, resulting in increasing continuum bandwidth with each
observation. All observations were obtained with a mean frequency
between 341 and 343 GHz (870 $\mu$m).

HERS1 was first observed in the SMA subcompact configuration (maximum
baselines $\sim$ 45 m) on 2016 September 20  (PID: 2016A-S007; PI: Cooray). The weather was good and
stable, with a mean $\tau_{\rm 225\ GHz}$ ranging from 0.07 to 0.09 (translating to
1.2 mm to 1.6 mm precipitable water vapor(pwv)). The target
observations were interleaved over a roughly 7.2 hr transit period,
resulting in 190.5 minutes of on-source integration time for
HERS1. Observations of HERS1 were next obtained in the SMA extended
configuration (maximum baselines $\sim$ 220 m) on 2016 October 16 . The
weather was good, with a mean $\tau_{\rm 225\ GHz}$ of 0.04 to 0.06 (0.6 mm to 1.0 mm pwv). The target observations were interleaved over a roughly 9.5 hr transit period, resulting in 225 minutes of on-source integration time for HERS1. The last observations of HERS1 were conducted in the SMA very extended configuration (maximum baselines $\sim$509 m) on 2017  September 30 and again on October 6 (PID: 2017A-S036; PI: Cooray). For the September 30 observation, only six antennas were available due to a cryogenics issue, which recovered in time for the October 6 observations. For the first observation, the
weather was very good, with a mean $\tau_{\rm 225\ GHz}$ of 0.06 (1.0 mm pwv), and phase stability was generally
very good.  In the second observation, the weather was somewhat worse,
with $\tau_{\rm 225\ GHz}$ rising from 0.06 to 0.085 (1 to 1.5 mm pwv), with
somewhat marginal phase stability. For both tracks, the target scans were
interleaved over a roughly 9 hr transit period, resulting in 160 minutes
(30 September) and 206 minutes (6 October) of on-source integration time for
HERS1. For all the observations, passband calibration was obtained using a
observations of 3C454.3, and gain calibration relied on periodic
observations of J0224+069. The absolute flux scale was determined from
observations of Uranus.

The integrated continuum visibility data for all four tracks were
jointly imaged and deconvolved using the Astronomical Image Processing
System (AIPS).  Using natural weighting of the visibilities, the
synthesized resolution is 580 mas $\times$ 325 mas (PA $27^{\circ}.2$ ), and the
achieved rms in the combined data map is 305 mJy\ beam$^{-1}$.

\subsection{ALMA observation}
We obtained four CO emission lines with ALMA from two programs. The CO(6-5) and CO(9-8) were obtained in project 2018.1.00922.S (PI: Riechers) on 2018 October 21 and November 30. Each execution used four spectral windows (SPW) covering the target lines. Each SPW was 2.000 GHz wide with 128 channels. Both observations were performed with ALMA 7 m antennas with a maximum baseline of 48.9 m, yielding the synthesis beam size 8$^{\prime \prime}$.14 $\times$ 5$^{\prime \prime}$.12 (PA= $76^{\circ}$) for CO(6-5) and 5$^{\prime \prime}$.28 $\times$ 3$^{\prime \prime}$.32 for CO(9-8) (PA= $88^{\circ}$). The bandpass and flux calibrators were J0237+2848 (CO(6-5)) and J0238+1636 (CO(9-8)). J0217+0014 was used as phase calibrator for both executions. The integration time was 18 minutes and 15 minutes reaching the rms of 1.4 mJy beam$^{-1}$ over the 266 MHz bandwidth (CO(6-5)) and 1.2 mJy beam$^{-1}$ over the 388 MHz bandwidth (CO(9-8)) respectively. CO(7-6) and CO(5-4) were observed in another project, 2016.2.00105S (PI: Riechers).
The observation covered the four frequency ranges of 211.06-214.90 GHz, 226.12-229.98 GHz, 147.02-150.88 GHz, and 159.15-162.99 GHz. Data were acquired on 2017 September 09 and 21 using ALMA 7 m antennas with a total on-source integration time of 11.8 and 10.3 minutes, respectively. Calibrators used for bandpass, flux, and phase calibrations were J0006-0623, Uranus and J0217+0144. The rms of the data reach 1.040 mJy beam$^{-1}$ over bandwidth 310.6 MHz for CO(7-6) and 1.800 mJy beam$^{-1}$ over the 221.6 MHz bandwidth for CO(5-4). The synthesis beam were 6$^{\prime \prime}$.88 $\times$ 4$^{\prime \prime}$.64 at a PA of $87^{\circ}$ (CO(7-6)) and 10$^{\prime \prime}$.50 $\times$ 5$^{\prime \prime}$.90 at a PA of $80^{\circ}$ (CO(5-4)). The CI fine-structure line (refer as CI(2-1) hereafter) was also detected in the same detection window of CO(7-6) at the frequency $\nu_{obs}= 227.8$ GHz. All data were mapped using {\it tclean} task in the {\sc casa} package (v.5.1.1 or 5.4.0) with a natural weighting. Table \ref{tab:observation} lists a brief summary of ALMA observations.

\subsection{SOFIA/HAWC+ observation}

HERS1 was observed with the HAWC+ instrument \citep{Dowell2013} on board SOFIA on 2017 November 15 under the Cycle 5 program PID 05-0087 (PI: Cooray), with results of a similar study with HAWC+ from the same program reported in \cite{Ma2018}. HAWC+ is capable of carrying out far-infrared imaging in five bands from 40 to 300 $\mu$m. We obtained the observation in Band C at 89 $\mu$m with a bandwidth of 17 $\mu$m in the Total-Intensity OTFMAP configuration. The HAWC+ Band C image has a field of view of $4.2^\prime \times 2.7^\prime$ with a PSF (FWHM) of 7$^{\prime \prime}$.8. The total effective on-source time is about 5250 s. The raw data were processed through the CRUSH pipeline v2.34-4 \citep{Kovacs2008}, and the final Level 3 data product was flux calibrated at the SOFIA Science Center. The flux calibration error is about 10\%. The resulting map has an rms noise level of $\sim$ 30 mJy beam$^{-1}$. The imaging data do not show a clear detection of HERS1 at its location. The 3$\sigma$ upper limit, extracted for a point source with the HAWC+ Band C PSF at the peak location of HERS1, is 168 mJy and is shown in Figure \ref{fig:SED}.

\subsection{Archival  observations }
HERS1 was observed by extensive programs covering different bands of emission lines and continua as mentioned above. Here we present a brief summary of the archival data of previous observations that were used in this paper.

\cite{Geach2018} performed an observation on 2017 December 11 and 14  (PID: 2017.1.00814.S; PI: Ivison) using the ALMA 12m array. The frequency ranges were 126.26-130.01 GHz and 138.26-142.01 GHz. A 1$\sigma$ rms sensitivity of 150 $\mu $Jy per 23 MHz channel was acquired with a total of 1.8 hr of on-source integration. The resulting beam size is $0^{\prime \prime}.27$ $\times$ $0^{\prime \prime}.21$. The data resulted in the discovery of three emission lines, CN(4-3) at 127.6 GHz, CO(4-3) at 129.8 GHz, and CI(1-0) at 138.5 GHz (for details see their paper). The NII(1-0) fine-structure line was also observed by this project at 441.2 GHz on 2018 August 26. A rms noise of 3 mJy \ beam$^{-1}$ in a 15.6 MHz channel was reached after a 42 minutes of total on-source integration. The synthesis beam is 0$^{\prime \prime}$.32 $\times$ 0$^{\prime \prime}$.25 (Table \ref{tab:observation}).  

The CO(3-2) line ($\nu _{\rm obs}$ = 97.3 GHz) was detected on 2017 December 26 (PID:2017.1.01214; PI: Yun) with the 12 m ALMA array. The line was covered in the spectral window centered on the frequency of 97.308 GHz with 480 channels. The resulting synthesized beam size was $0^{\prime \prime}.45$ $\times$ $0^{\prime \prime}.32$ and the rms was 420 $\mu$Jy \ beam$^{-1}$ with a bandwidth 3.9 MHz (Table \ref{tab:observation}).

Apart from the WFC3 F125W band, the lensing galaxy has been observed in two other HST bands; these are HST/WFC3 F110W (PID: 15242; PI: Marchetti) and HST/WFC3 F160W (PID: 14653; PI: Lowenthal). Both photometry results were included in the SED fitting. In the SED fitting, the following observations were also adopted: Spitzer/IRAC 3.6\ $\mu$m and 4.5\ $\mu$m (PID: 14321; PI: Yan), Herschel/SPIRE 250\ $\mu$m, 350\ $\mu$m, and 500\ $\mu$m \citep{Viero2014}, SCUBA 850\ $\mu$m, AzTEC 1.1 mm \citep{Geach2015}, IRAM 1.3 mm, and ACT 2.026 mm \citep{su2017}.

\section{Lens Model}
\label{section:model}

\subsection{HERS1}

HERS1 is a gravitationally lensed galaxy magnified by two foreground galaxies (a primary galaxy G1 and a satellite G2 located at the northwest of G1 shown in Figure \ref{fig:Fig three color})  at the same redshift $z=0.202$. In order to derive its intrinsic properties, we first built a lens model of this system.

\begin{figure*}[]
\centering
\includegraphics[trim=0cm -1cm 0cm 0cm, scale=0.55]{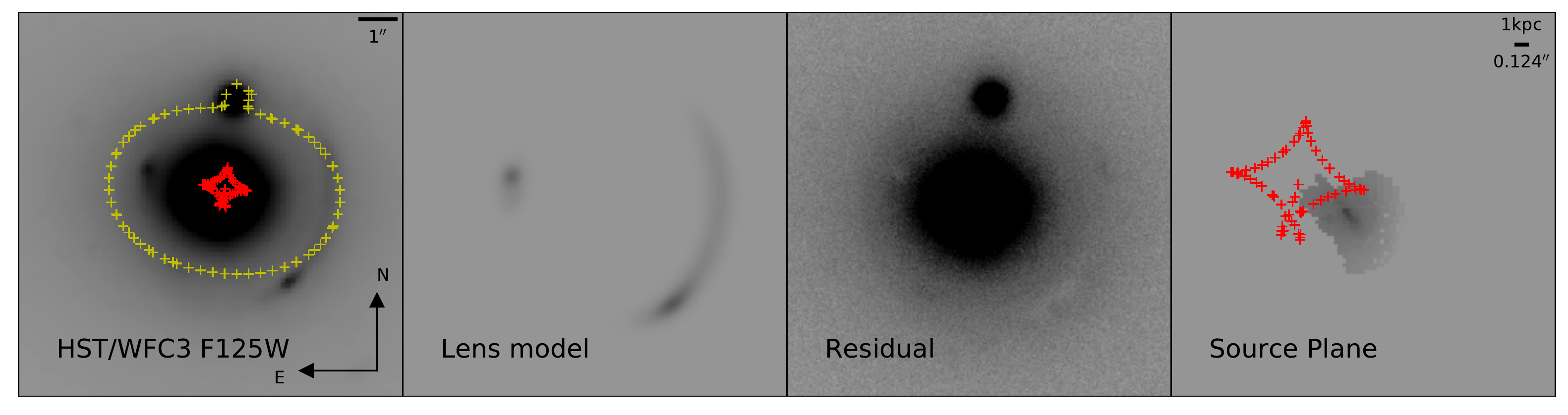}
\caption{Lens-modeling results of HERS1. First column: HST/WFC3 F125W-band image of HERS1 with two foreground galaxies and a partial Einstein ring observed. The yellow and red cross marks are the critical and caustic lines derived from the best-fit model. Second column: the reconstructed lens-plane image of the best-fit model results . Third column: the residual map after subtracting the model from the original image. Last column: the source-plane reconstruction along with the caustic line.}
\label{fig:Fig model}
\end{figure*}

\begin{figure*}[]
\centering
\includegraphics[width=0.8\textwidth, scale=1]{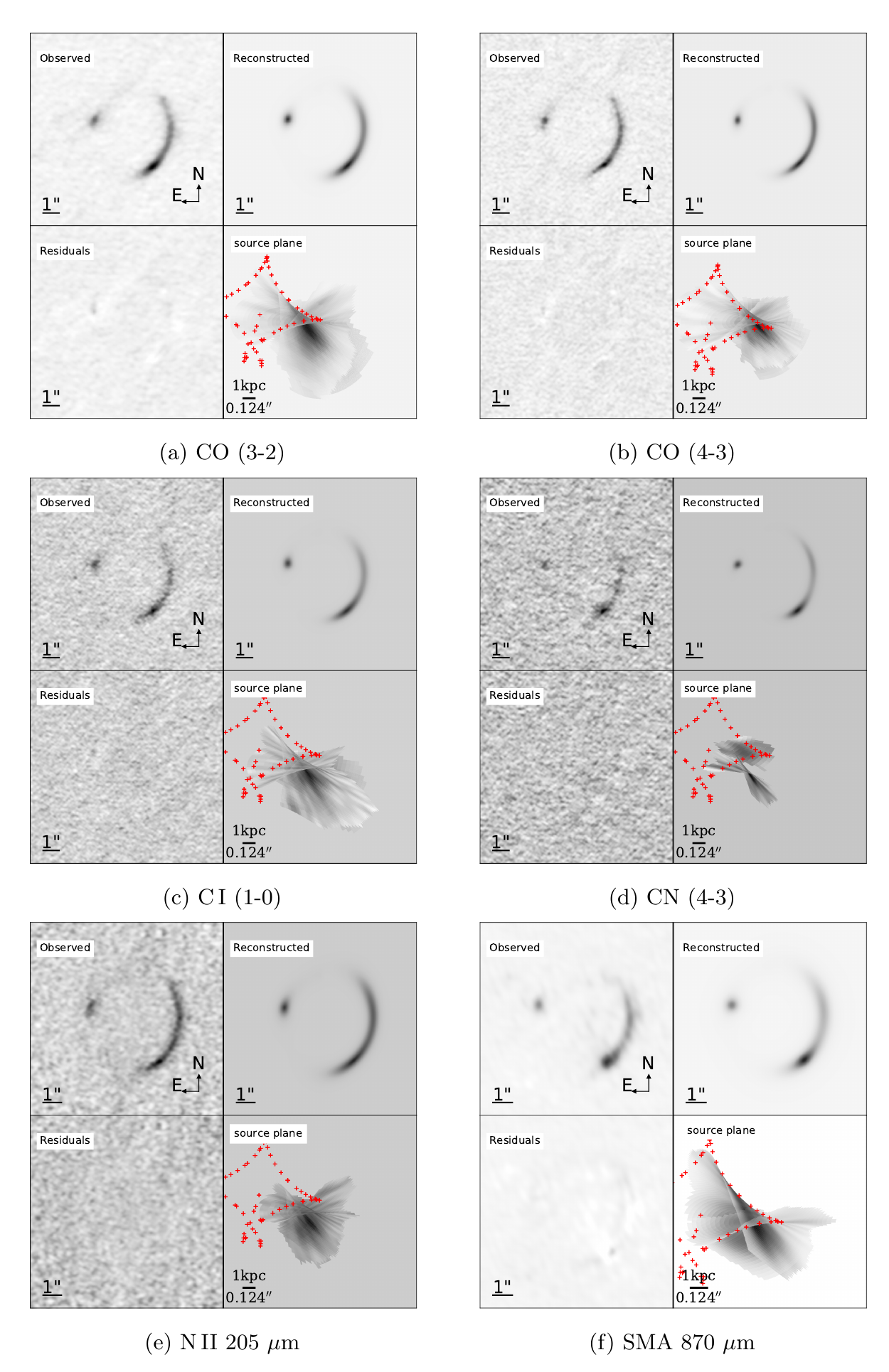}
\caption{Observation of emission lines from ALMA and SMA $870\ \mu$m along along with modeling results using the best-fit model derived from the high-resolution HST image. Labels (a)-(f) list the line species. The channel-dependent source plane reconstructions is available in the online version of the paper.}
\label{fig:Fig lines}
\end{figure*}

We used the publicly available code {\sc lenstool}\footnote{\url{http://projets.lam.fr/projects/lenstool/wiki}} to construct the best-fit model. The best-fit parameters were found by performing a Bayesian Markov Chain Monte Carlo (MCMC) sampling. The lensing system is mainly made up of two parts as shown in Figure \ref{fig:Fig three color}. We used the HST/WFC3 F125W high-resolution image to constrain the lens model parameters. We first fitted the light profile of the lensing galaxies (G1 and G2) by a S$\acute{\rm e}$rsic function using {\sc galfit} \citep{Peng2010}. We then subtracted the modeled foreground galaxies to obtain the lensed image. We checked the robustness of the {\sc galfit} results by performing a photometric decomposition with the {\sc gasp2d} code \citep{Mendez2008,Mendez2017}. Therefore, we kept the simpler model defined by a single S$\acute{\rm e}$rsic function. The lensed components were then identified using the {\sc photutils} \footnote{\url{https://photutils.readthedocs.io/en/stable/}} package \citep{Bradley2020} and broken into four ellipses. The elliptical size and fluxes of these ellipses were then measured as input information for {\sc lenstool}. We chose a singular isothermal ellipsoid profile for the primary galaxy G1 and a singular isothermal sphere profile for G2. We fixed the positions of the galaxies, so the model was parameterized by the ellipticity $e$, the position angle $\theta$ and the velocity dispersion $\sigma_1$ for G1 and velocity dispersion $\sigma_2$ for G2. The redshift of foreground galaxies and background galaxies was fixed at $z=0.202$ and $z=2.553$, respectively. The optimization output provided the best-fit results. Table \ref{Table:lens} gives the best-fit parameters of the lensing galaxies. The reconstructed images obtained from the best-fit model are shown in Figure \ref{fig:Fig model}. From this model, we obtain the luminosity-weighted magnification factor to be $\mu_{\rm star} = 13.6\pm 0.4$. The best-fit model established above was also used to reconstruct images of the aforementioned high-resolution emission lines and SMA dust continuum as shown in Figure  \ref{fig:Fig lines}. The dust map provides a magnification factor $\mu_{\rm dust}=12.8\pm 0.3$. The source-plane images are reconstructed using the {\sc cleanlens} algorithm within {\sc lenstool}, the results are also illustrated in Figure \ref{fig:Fig model} and Figure \ref{fig:Fig lines}. Our model is generally consistent with the model of \cite{Geach2015} while our ellipticity for G1 (0.34) is larger than their results (0.12). The magnification of the stellar and dust components is also similar to that of other bands \citep{Geach2015,Geach2018,Rivera2019} that $\mu=$11 - 15.

\subsection{Second lensed source}
\begin{figure}[h]
\centering
\includegraphics[scale=0.6]{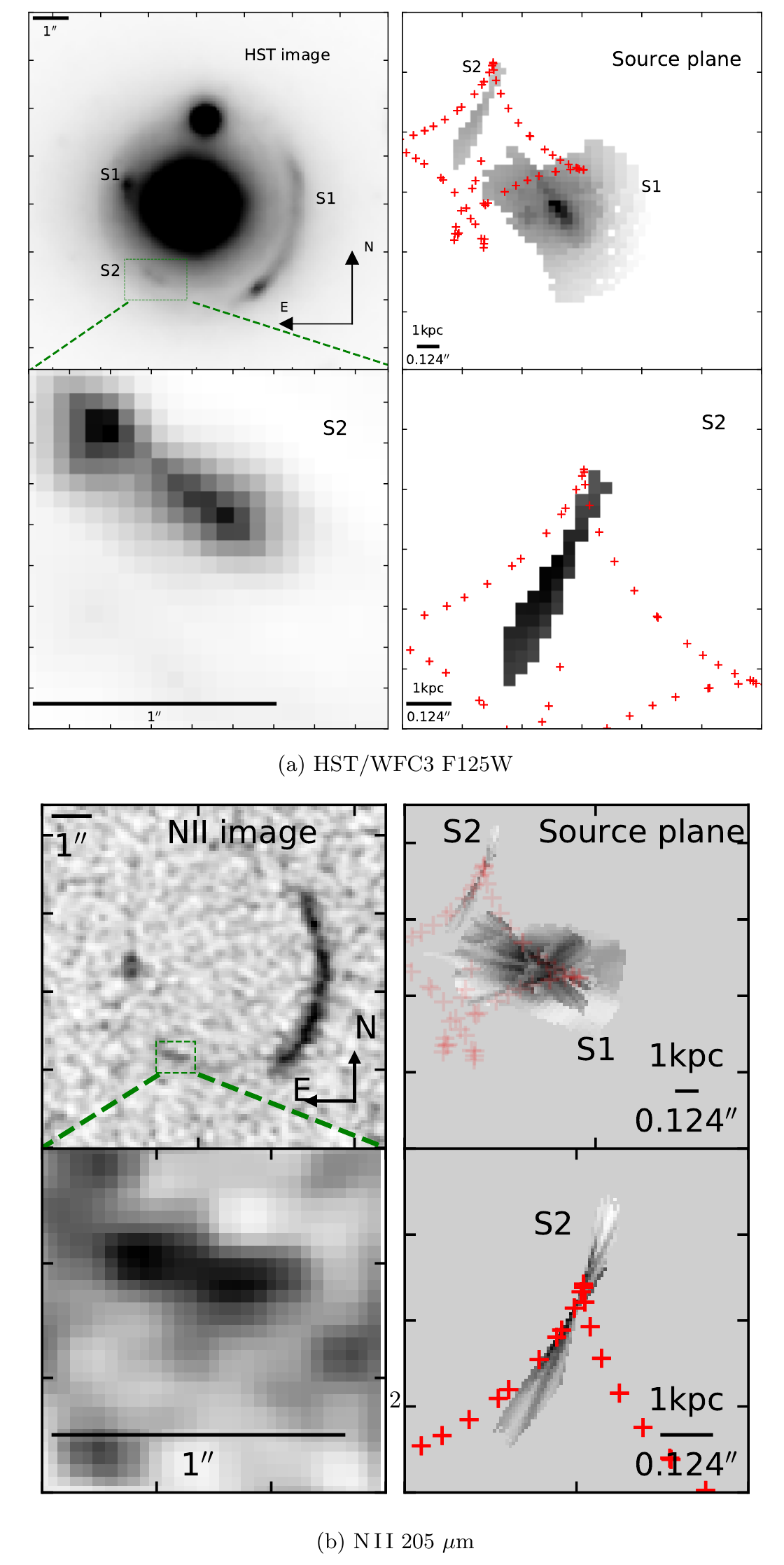}
\caption{Panel (a), left: HST/WFC3 F125W image. The upper panel shows the entire image of the lensed arcs. The green box displays the position of the additional clump. A zoomed-in image of this part is shown in the lower panel. Right: source-plane reconstructions with the caustic line overplotted in red using the best-fit model. The upper panel shows the reconstruction of all lensed arcs with S1 corresponding to HERS1 and S2 corresponding to the additional clump. The lower panel shows a zoomed-in image of S2. Panel (b): NII 205 $\mu$m images in the same order as panel (a).}
\label{fig:Fig source e}
\end{figure}

\begin{figure}[h]
\centering
\includegraphics[width=0.4\textwidth, scale=0.5]{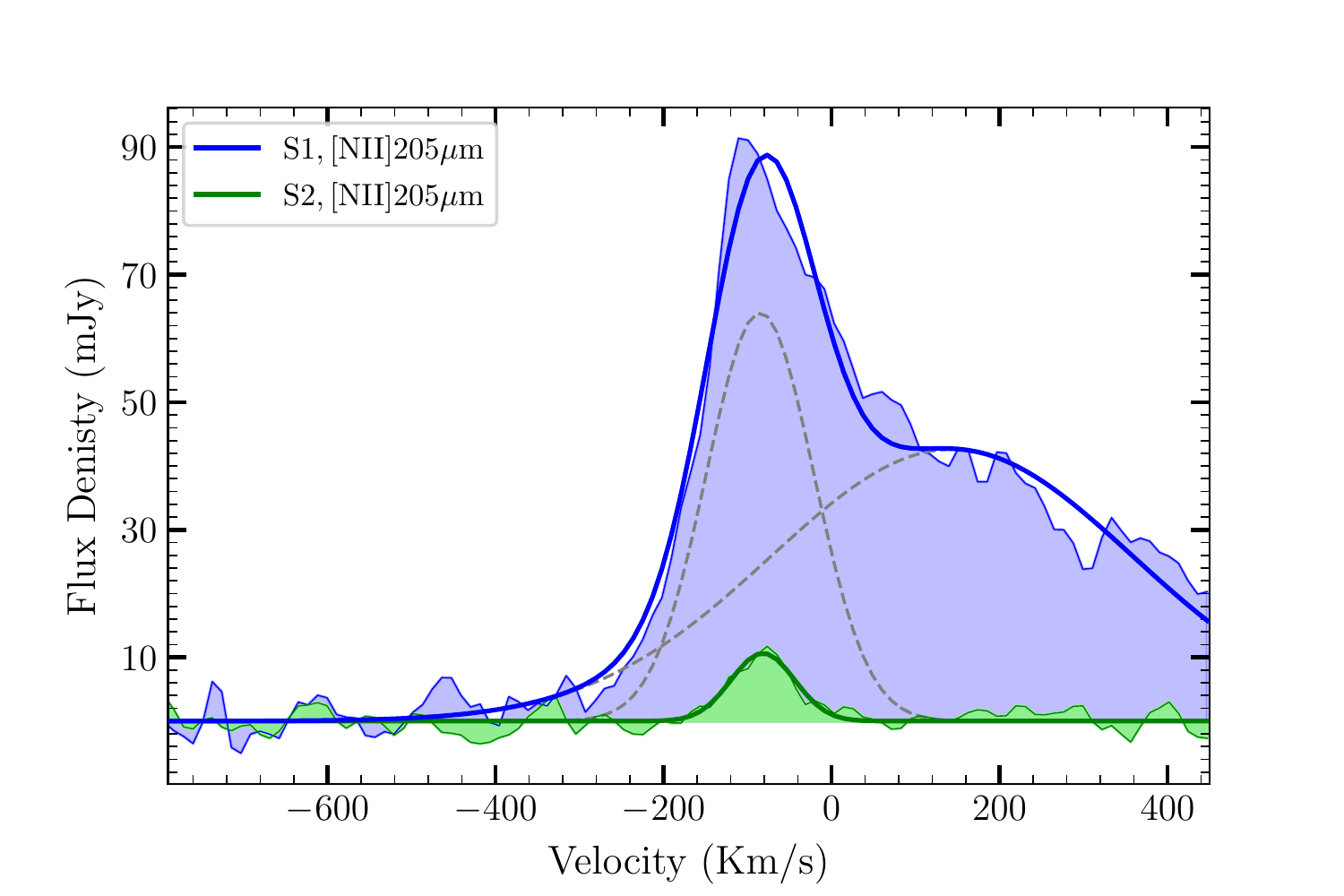}
\caption{Observed [N\,II] 205 $\mu$m spectral profile of HERS1 and the additional clump for comparison. The blue component shows the line profile of HERS1, and the solid blue line demonstrates the double-Gaussian fit result with the individual components overlaid as grey dashed lines. The green component corresponds to the profile of the additional clump. A single Gaussian function was used, as shown in the solid green line.}
\label{fig:Fig NII profile}
\end{figure}

In the high-resolution HST data, an additional clump is detected, as shown in Figure \ref{fig:Fig source e}. Our model does not have a predicted counterimage in that position. We traced this part to the source plane using the best-fit model above; the corresponding source located in the northeast of the main source as shown in Figure \ref{fig:Fig source e} which suggests it is an extra individual component. S2 is also detected in ALMA [NII] 205 $\mu$m (data from \citealt{Geach2018} and described in \citealt{Doherty2020}). While this study failed to identify the detection with a second source, we confirm the presence of S2 at the same redshift as S1 through a combination of HST/WFC3 and [NII] observations. The line profiles of S1 and S2 are compared in Figure \ref{fig:Fig NII profile}. The extra component has a narrower line width, which also implies that it comes from a different region. The integrated [NII] 205 $\mu$m spectral line flux density of S1 and S2 are $35.72 \pm 1.35$ Jy km s$^{-1}$ and $1.05\pm 0.62$ Jy km s$^{-1}$, respectively.  

We also made a comparison of the ratio of the detection results. These are listed in Table \ref{Table:Ratios of S1 and S2}. For the nondetection of the SMA dust continuum of S2, we adopted a $3\sigma$ upper limit when calculating the expected ratios. As shown in the table, the new component contains much less dust compared to S1, but the [NII] detection shows a higher ionization fraction. Assuming a low electron density (below 44 cm$^{-3}$), the [NII] 205 $\mu$m emission yields a lower limit on the star formation rate of S2 of $\sim$ 3 $M_{\odot}$ yr$^{-1}$ \citep{Herrera2016,Harrington2019}. These results are more consistent with a normal star-forming galaxy that is likely merging with the central source (S1) of HERS1. As the detections of S2 are limited to rest-frame optical emission in HST/WFC3 and [NII] 205 $\mu$m with ALMA, it is difficult to determine its  exact nature. Further deeper observations such as the CO emission and continuum flux densities in the optical to near-IR rest-frame wavelengths are needed to extract gas and stellar properties and to establish its physical properties.

\begin{table}[h!]
    \caption{Lens-modelling results and surface brightness model for lensing galaxies }
    \label{Table:lens}
    \centering
    \begin{tabular}{cccc}
    \hline
    \hline \multicolumn{4}{c}{Lensing model}  \\
    \hline Object   &   Quality     &   Value           &   Units       \\
    \hline G1       &   $e$           &   $0.34\pm 0.06$  &   ...\\
           ...      &   $\theta$      &   $-8\pm2$        &   deg\\ 
           ...      &   $\sigma_1$      &   $318\pm3$       &   km\ s$^{-1}$\\
           ...      &   RA          &   02:09:41.27     & h:m:s    \\  
           ...      &   Dec         &   +00:15:58.53    & d:m:s    \\
           ...      &   $z$           &   0.202           & ...      \\
           G2       &   $\sigma_2$      &   $66\pm10$       &   km\ s$^{-1}$\\
           ...      &   RA          &   02:09:41.24     & h:m:s    \\  
           ...      &   Dec         &   +00:16:00.84    & d:m:s    \\
           ...      &   $z$           &   0.202           & ...    \\
    \hline \multicolumn{4}{c}{Surface brightness model}  \\
    \hline G1       &  $R_{\rm e}$        &   2.3              & arcsec \\
           ...      &  $n$            &   3.2             & ...   \\
           ...      &  $b/a$          &   0.93            & ...   \\
           ...      &  PA           &   119             & deg   \\
           G2       &  $R_{\rm e}$        &   0.4               & arcsec \\
           ...      &  $n$            &   1.3             & ...   \\
           ...      &  $b/a$          &   0.95            & ...   \\
           ...      &  PA           &   118             & deg   \\
    \hline
    \end{tabular}
\end{table}

\begin{table}[h!]
    \caption{Flux ratios of HST/WFC3 F125W, [NII] 205 $\mu$m and SMA 870$\mu$m between S1 and S2.}
    \label{Table:Ratios of S1 and S2}
    \centering
    \begin{tabular}{ccc}
    \hline
    \hline Ratio           &   S1                    &   S2          \\
    \hline [NII]/HST\tablenotemark{a}       &   $2001.69\pm 140.36$   &   $1195.31 \pm 753.17$ \\  
    \hline SMA/[NII]      &   $3.37 \pm 0.14    $   &   $<1.60$ \tablenotemark{b}               \\
    \hline SMA/HST         &   $6734.01 \pm 416.39$  &   $ <1909.37$\tablenotemark{b}          \\
    \hline
    \end{tabular}
    \tablenotetext{a}{The  total integrated [NII] line flux densities from the ALMA spectral line cubes at rest-frame 205 $\mu$m were extracted for S1 and S2 (Section 3.2) in units of millijansky. The HST and SMA flux densities are also converted to units of millijansky so the ratio is unitless. The HST value is for the observed F125W filter while the SMA flux density is for the 870 $\mu$m band.}
    \tablenotetext{b}{The undetected SMA flux density for S2 (in millijansky) was calculated by adopting a $3\sigma$ upper limit.}
\end{table}

\section{Physical Properties}
\label{section:physical}

\subsection{CO line properties }

\begin{figure}
\centering
\includegraphics[trim=0cm 0cm 0cm 0cm, scale=0.6]{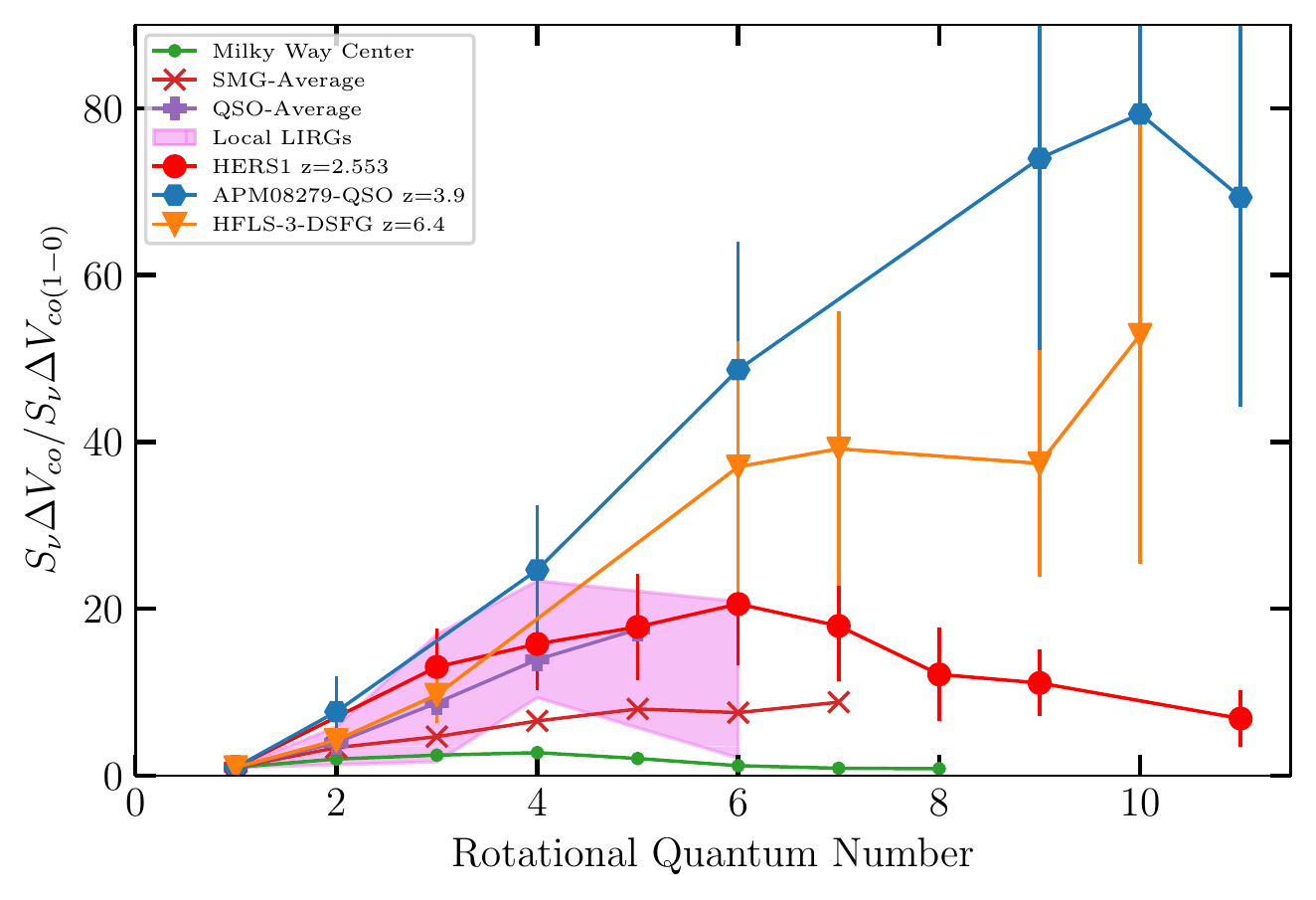}
\caption{Observed CO SLEDs normalized by CO(1-0) flux density with other well-studied systems for comparison.  }
\label{fig:Fig CO SLED}
\end{figure}

CO SLED is an effective tool that can be used to reveal the bulk physical properties of a galaxy. HERS1 has multiple CO line detections up to $J_{\rm up}=11$. We first calculated the observed CO SLED combining the data in this paper and the results from the literature \citep{Geach2018, Harrington2021}. Figure \ref{fig:Fig CO SLED} shows the line variations normalized by CO(1-0). Other well-studied systems are also shown for comparison \citep{Fixsen1999,Papadopoulos2012,Bothwell2013,carilli2013,Riechers2013}.
Overall, the CO flux increases with $J_{\rm up}$ and peaks at $J_{\rm up}=6$, then declines toward higher rotation numbers. The excitation is higher than the average of SMGs but not as high as the extremely excited galaxies such as APM 08279 and HFLS-3. The ladder shape is also similar to some local ULIRGs and starburst galaxies \citep{Mashian2015,Rosenberg2015}.

\begin{figure}
\centering
\includegraphics[trim=0cm 0cm 0cm 0cm, scale=0.65]{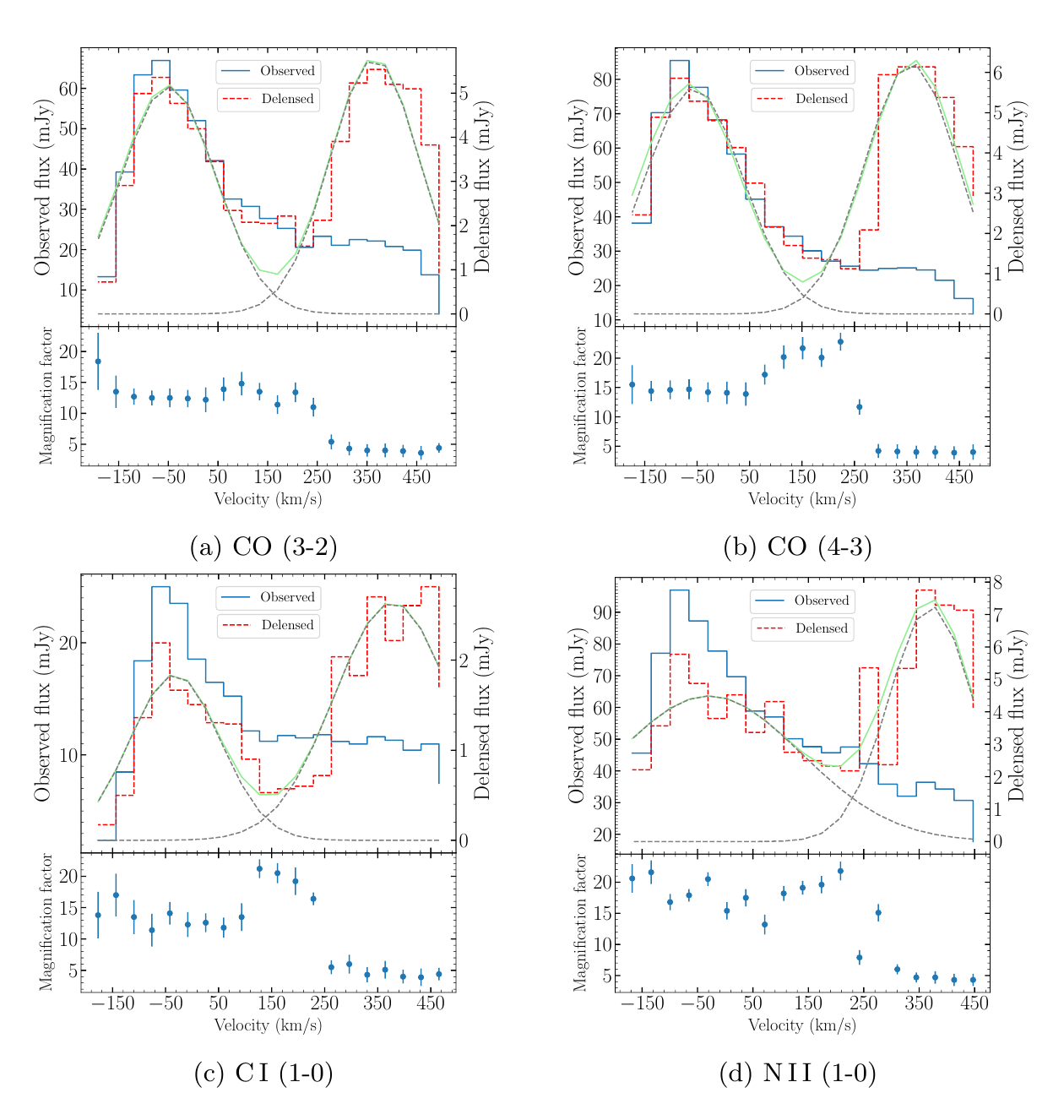}
\caption{(a)-(d): Spectral lines from observations with high-resolution lensed images. In each subpanel, the upper component shows the delensed (dashed red line) and  observed (solid blue line) spectra. The delensed line was fitted with a double-Gaussian model by fixing the FWHM of each component and the width between two peaks. The best-fit result is shown as a solid green curve while the individual components are overlaid in dashed gray. The lower panels show the magnification factor for each velocity channel.}
\label{fig:Fig delensed}
\end{figure}

\begin{figure}
\centering
\includegraphics[trim=0cm 0cm 0cm 0cm, scale=0.85]{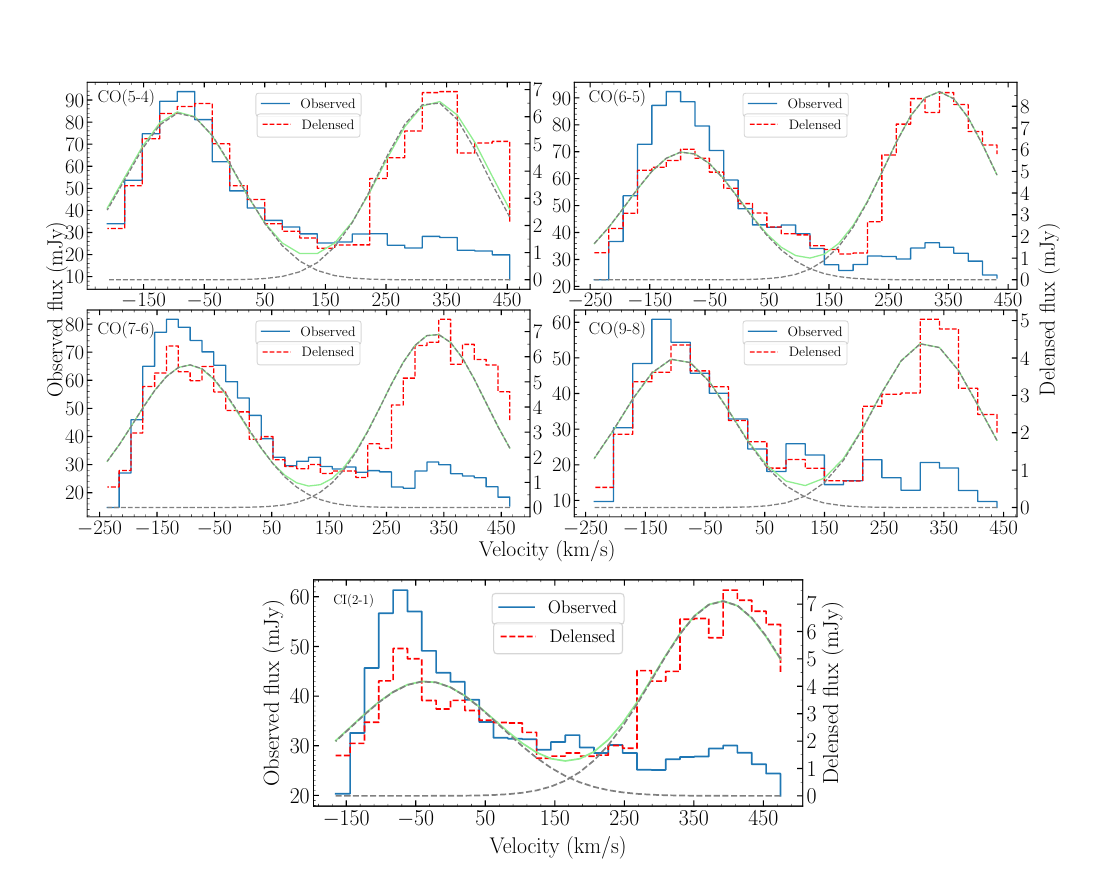}
\caption{Same as Figure \ref{fig:Fig delensed}, but showing the high-$J_{\rm up}$ CO and C\,I lines. The magnification factor is taken from CO(4-3) for CO lines and CI (1-0) for the CI (2-1) line.}
\label{fig:Fig highco}
\end{figure}

In order to further study the intrinsic properties of the source, we applied the magnification correction to these line intensities. Due to the lack of data of some lines, we only calculated the delensed fluxes of lines $J_{\rm up}=3,4,5,6,7,9$. For lines CO(3-2) and CO(4-3), we first rebinned the data to a width of 36 km\ s$^{-1}$ and calculated the magnification factor on each channel map based on the best-fit lens model and high-resolution lensed image. The results are shown in Figure \ref{fig:Fig delensed} (the CI and [NII] lines are presented as well using the same method). For lines with $J_{\rm up} \geq 5$, we adopted the magnification factor of CO(4-3) velocity channels due to the lack of high-resolution images. It is suggested that differential lensing may cause a bias on the CO SLED calculation. \cite{Serjeant2012} shows the $1\sigma$ dispersion of CO(6-5) can reach $20\%$ of the mean valve for 500 $\mu$m-selected sources with $\mu > 10$. Frequency-dependent lens models are required to verify our assumption, but this is beyond the scope of this paper. Though differential lensing distorts the CO SLED, the mid-$J$ lines may have similar distortion. The line widths of our CO lines are close to each other, which implies they arise from similar emission regions and this is consistent with the assumption. So we consider using the magnification factor of CO(4-3) as the factor of higher-$J_{\rm up}$ lines is a moderate assumption.   Figure \ref{fig:Fig highco} shows the observed and delensed line profiles.   

\begin{figure}
\centering
\includegraphics[trim=0cm 0cm 0cm 0cm, scale=0.6]{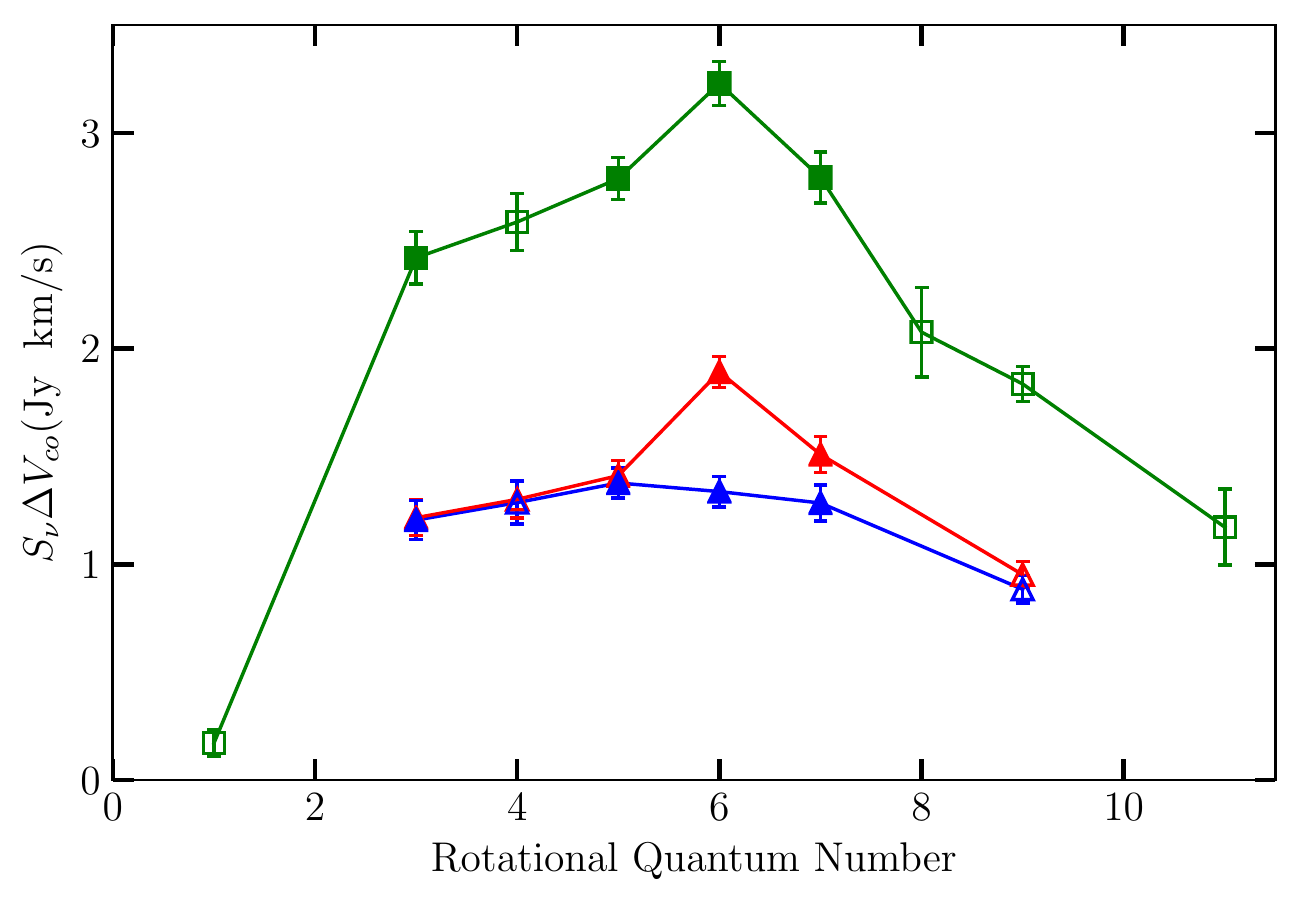}
\caption{Velocity-integrated line fluxes corrected for magnification. Green symbols represent the total flux densities. Data points plotted with open squares were described in \cite{Geach2018}, \cite{Harrington2021}, and \cite{Riechers2021}. Red and blue symbols show the flux densities of the red and blue kinematic components, respectively.}
\label{fig:Fig iladders}
\end{figure}

The observed line fluxes present a double-horned profile; this feature becomes more obvious after correcting the magnification in each velocity channel. We also separated the line profile into two individual components, a `blue' part and a `red' part based on their velocity. A double-Gaussian model was used to fit each delensed line by fixing the peak separation ($\Delta v =419\ \rm{km\ s^{-1}}$) and the two FWHMs ($\rm{FWHM_b}=221\  \rm{km\ s^{-1}}$, $\rm{FWHM_r}=209\ \rm{km\ s^{-1}}$). The model lines were also illustrated in Figure \ref{fig:Fig delensed} and Figure \ref{fig:Fig highco}. Velocity-integrated flux densities were then measured from the Gaussian functions. Results are reported in Table \ref{tab:lines}. The intrinsic SLED of the total source as well as the individual components was also shown in  Figure \ref{fig:Fig iladders}. As we can see, the magnification factors vary with the velocity. The blue components have higher magnification than the red components, so the delensed flux profiles show more symmetrical forms compared to the observed spectra.

To further investigate the molecular gas, we apply the {\sc lvg} code to the CO SLED. We adopt the code {\sc radex} \citep{vandertak2007} to fit our CO fluxes. {\sc radex} is a non-LTE analysis code to compute the atomic and molecular line intensities with an escape probability of $\beta = (1-e^\tau)/\tau$. CO collision files are taken from the Leiden Atomic and Molecular Database (LAMDA) \citep{schoier2005}.

The input parameters of {\sc radex} include the molecular gas kinetic temperature $T_{\rm k}$, the volume density of molecular hydrogen $n_{\rm H_2}$, the column density of CO, $N_{\rm CO}$, and the solid angle of the source. The velocity gradient is fixed to 1 km\ s$^{-1}$, so the CO column density $N_{\rm CO}$ also equals to the column density per unit velocity gradient $N_{\rm CO}/dv$. We are only concerned with the first three parameters, $T_{\rm k}$, $n_{\rm H_2}$, and $N_{\rm CO}/dv$, because the resulting CO SLED shape is not dependent on the solid angle. Instead of using {\sc radex} grids, which produce a grid of CO emission fluxes given a range of parameters, we adopted a Bayesian method and performed an MCMC calculation to fit {\sc radex} results with the observed fluxes. This allows a faster convergence and a better sampling in the parameter space \citep{yang2017}. The code {\sc radex\_emcee} was used for the calculation \footnote{\url{https://github.com/yangcht/radex\_emcee}}. This package combines {\sc pyradex}\footnote{\url{https://github.com/keflavich/pyradex}}, a {\sc python} version of {\sc Radex} converted by Ginsburg, and {\sc emcee}\footnote{\url{ https://github.com/dfm/emcee}} to achieve the fitting. 

Previous studies show that the CO emissions are likely dominated by two components \citep[e.g.,][]{Daddi2015,yang2017,Canameras2018} and a single-component model is inadequate for fitting our SLED. So we adopted a two-component model including a warmer high-excitation component and a cooler low-excitation component in our fitting as the one-component model poorly fitted the CO SLED. That model allows two sets of parameters for different physical conditions of the two components. We applied the flat log-prior for $n_{\rm H_2}$, $N_{\rm CO}/dv$ \citep[e.g.,][]{Spilker2014}. The ranges of the parameters  are taken as $n_{\rm H_2} = 10^2-10^7$ and $N_{\rm CO} = 10^{15.5}-10^{19.5}$ for both two components. An extra limit of $dv/dr$ with a range of $0.1-1000$ km\ s$^{-1}$ is adopted \citep[e.g.,][]{Tunnard2016}, this provided a limit of the ratio between $N_{\rm CO}/dv$ and $n_{\rm H_2} $. For the kinetic temperature, two different prior ranges were chosen. The warmer component also has a flat log-prior with the range $T_{\rm k} = T_{\rm CMB}-10^3$ K , where $T_{\rm CMB}$ is the CMB temperature at the redshift of the source. For the cooler part, we set an additional limit that the temperature was close to the temperature of cold dust. As discussed by \cite{Goldsmith2001}, the temperature of gas and dust couple well at high density, $N_{\rm H_2} \geq 10^{4.5} $cm$^{-3}$, but this relation is not satisfied when $n_{\rm H_2} \leq 10^{3.5} $cm$^{-3}$. So we took a normal distribution of the cooler component, which gave a reasonable guess in the range $T_{\rm CMB}-90$ K. Further,  the size of the cooler component was set to be larger than the warmer component, inspired by observations of the sizes of different CO line emission regions \citep[e.g.,][]{Ivison2011b}.

\begin{figure}
\centering
\includegraphics[trim=1cm 0cm 0cm 0cm, scale=0.85]{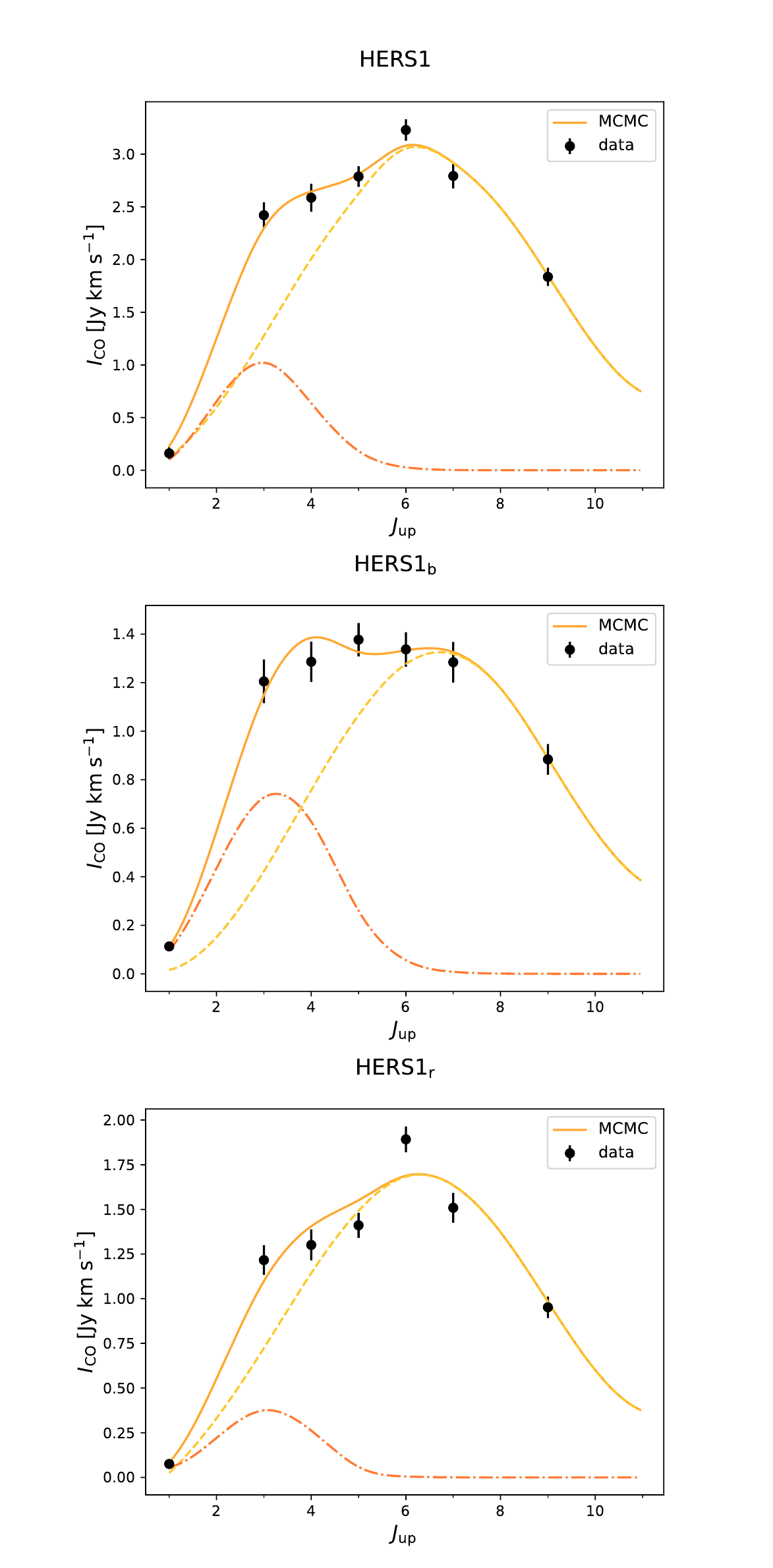}
\caption{Top: delensed CO SLED of HERS1 and the best-fit LVG model using two excitation components (solid orange line). The dotted-dashed red line and dashed orange line represent the lower- and higher-excitation components, respectively. Middle: same as the top panel but shows the result of the blue kinematic component. Bottom: same as the top panel but shows the result of the red kinematic component.}
\label{fig:Fig LVGSLED}
\end{figure}

\begin{figure*}
\centering
\includegraphics[width=0.78\textwidth, scale=0.4]{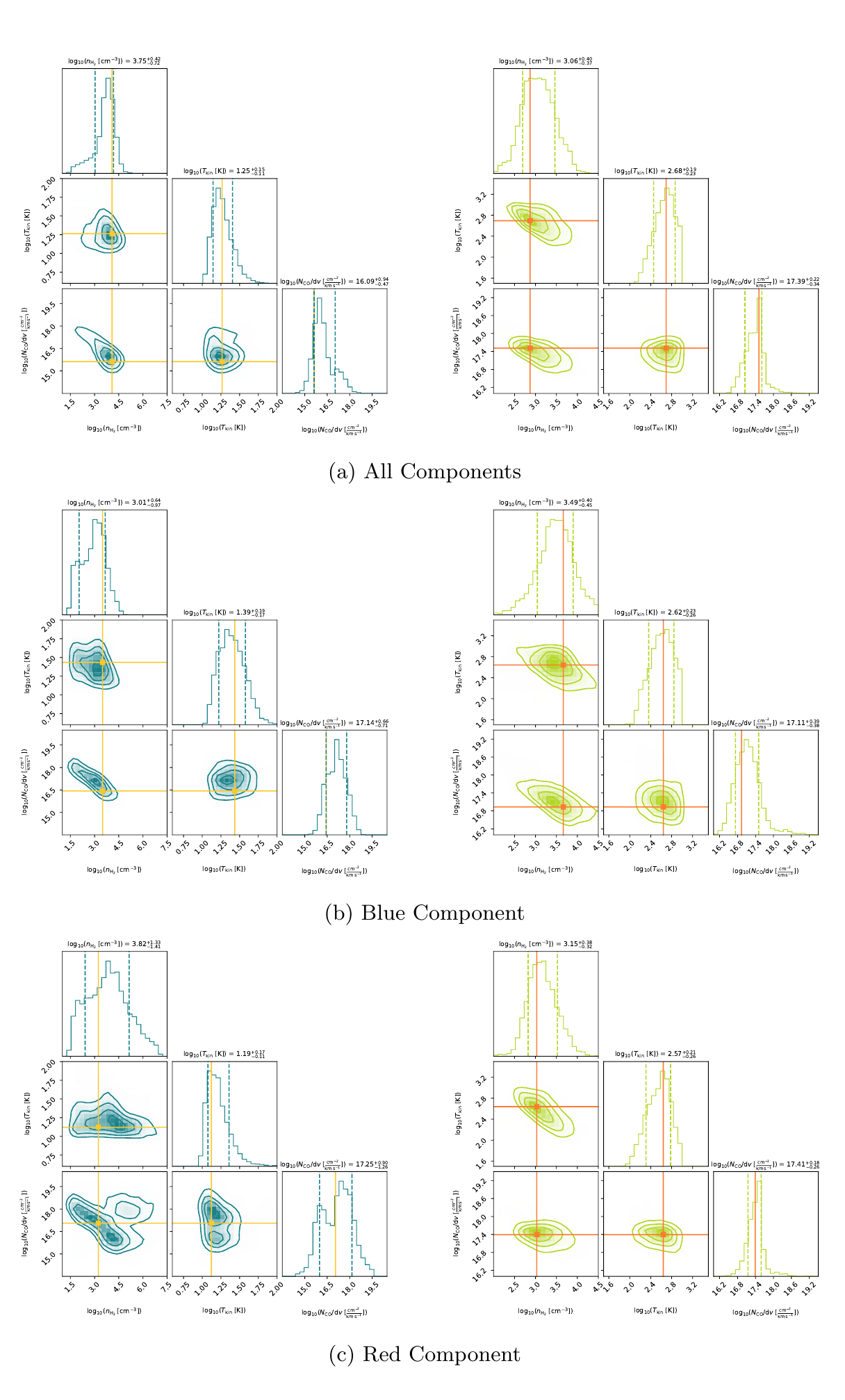}
\caption{Posterior probability distribution of $n_{H_2}$, $T_{\rm kin}$ and $N_{\rm CO}/dv$ derived from the MCMC sampling of the two-component model. Solid lines show the maximum posterior probability of each parameter, while the dashed lines show the $\pm1\sigma$ range. The results are also listed in each histogram. The left panel corresponds to the lower-excitation component, while the right panel corresponds to the higher-excitation component. From top to bottom are the results of the total SLED and the blue and red kinematic components, in the same order as Figure \ref{fig:Fig LVGSLED}.}
\label{fig:Fig LVGmod}
\end{figure*}

Figure \ref{fig:Fig LVGSLED}, Figure \ref{fig:Fig LVGmod} and Table \ref{tab:LVGresults} show the best-fitting two-component results. The results indicate that two components to the CO SLED can give a better description of the CO excitation, a low-excitation component with a cooler temperature and a high-excitation component with a warmer temperature. The low-excitation component peaks at $J_{\rm up}=3$ and only contributes to $J_{\rm up}\leq 4$ excitation. The high-excitation component dominates the CO emission and peaks at $J_{\rm up}=6$. The gas density and temperature are consistent with the results of other studies of high-redshift SMGs \citep[e.g.,][]{Spilker2014,yang2017}. The mid-$J$ lines such as CO(6-5) and CO(7-6) trace the molecular gas, which is related to the star formation. Several works have shown that mid-$J$ CO luminosity has a linear correlation with the infrared luminosity $L_{\rm IR}$ \citep[e.g.,][]{Greve2014,Rosenberg2015,Lu2017}. The warm component, which contributes nearly all CO flux at $J_{\rm up}>4$, is thought to be more closely related to the star formation activity. \cite{yang2017} also found a correlation between the thermal pressure $P_{\rm th}$ (defined as $P_{\rm th} \equiv n_{\rm H_2}\times T_{\rm k}$) and the star formation efficiency (defined as ${\rm SFE}\equiv {\rm SFR}/M_{\rm gas}$), which suggested a tight relation of star formation and the gas in the warmer component in high-redshift SMGs. The warmer component has a high kinetic temperature, $T_{\rm k} = 479$ K, which is much higher than the dust temperature, $T_{\rm dust} = 35$ K. 

\begin{figure}
\centering
\includegraphics[trim=0cm 0cm 0cm 0cm, scale=0.6]{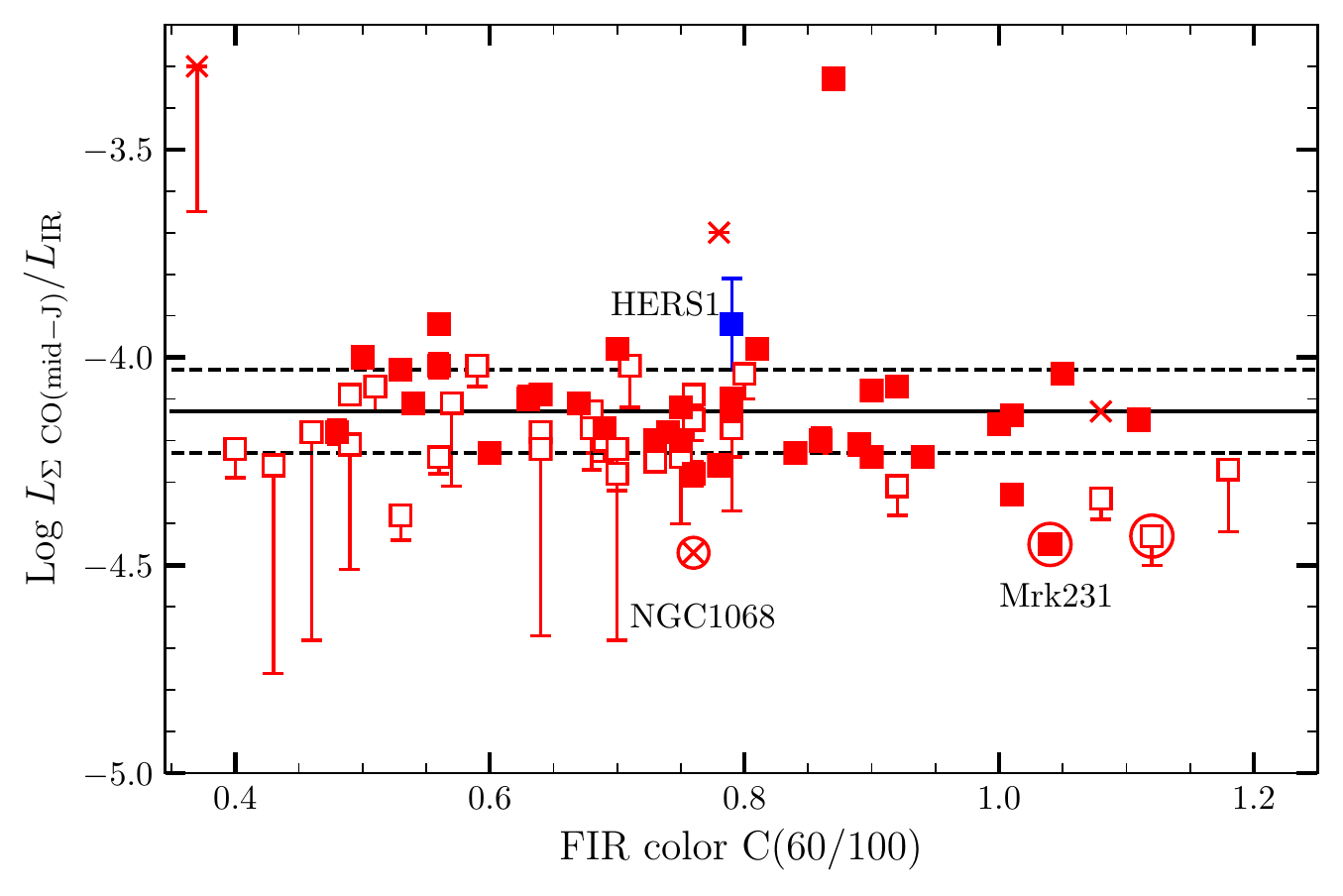}
\caption{The ratio of the combined luminosity of mid-$J$ lines (CO(5-4), CO(6-5), CO(7-6), CO(8-7), CO(10-9)) to $L_{\rm IR}$ as a function of the FIR color for HERS1 and the sample galaxies of \cite{Lu2014}. Galaxies with a significant AGN are illustrated with an extra open circle. The symbols are the same as \cite{Lu2014}.}
\label{fig:RmidJIR}
\end{figure}

Beacuse the excitation is dominated by the warmer component, the high-$T_{\rm k}/T_{\rm dust}$ ratio may imply extra heating mechanisms. Separately, it has also been suggested that HERS1 might contain an active galactic nucleus (AGN) \citep{Geach2015}. The AGN can be a strong heating source in galaxies as studied by a number of works \citep[e.g.,][]{Weiss2007,Gallerani2014}. The CO SLED shape provides simple diagnosis. \cite{Lu2014} has shown that the ratio of the combined luminosity of mid-$J$ lines to the infrared luminosity $L_{\rm IR}$ would be different in star formation (SF)-dominated and AGN-dominated galaxies. The SF-dominated galaxies have an average logarithmic ratio of -4.13, and the galaxies with a significant AGN contribution show a lower value. HERS1 has a ratio of $-3.92\pm 0.11$, suggesting the lack of a significant AGN impact on the CO SLED as shown in Figure \ref{fig:RmidJIR}. Two representative AGN-dominated galaxies, NGC 1068 and Mrk 231, suggest a much lower mid-$J$ to $L_{\rm IR}$ ratio. These two galaxies are also found to have higher CO emissions at $J_{\rm up}>10$. The high-$J$ CO lines can be employed to characterize the AGN heating as well. The ratio of high-$J$ to mid-$J$ CO line has been used to discriminate the starburst and AGN activity. We use the ratio of CO(10-9) and CO(6-5) to characterize the AGN heating, which has been used in the literature \citep[e.g.,][]{Wang2019,Jarugula2021}. An AGN can increase the high-$J$ excitation and will provide a flat SLED shape as well as a higher line ratio in high-$J$ and mid-$J$ lines. From the model results, we can get $L_{\rm CO(10-9)}/L_{\rm CO(6-5)}\sim 0.6-0.7$, which is similar to the Class II samples of \cite{Rosenberg2015}. It is suggested that most of the Class II galaxies have a low AGN contribution. The line ratio is also close to the average value of the local starburst galaxies as shown in \cite{Carniani2019}. Though we cannot completely rule out an AGN contribution, the line ratio result does not conclusively suggest AGN heating as the dominant mechanism. Further observations such as X-ray studies are needed to study the AGN in HERS1 in detail. Other heating sources such as mechanical processes invoked in some local starburst galaxies \citep[e.g.,][]{Hailey2008,Nikola2011} are also possible. The two individual kinematic components were then fitted in a similar manner. The two spectral components show similar excitation compositions and gas properties to the global SLED.

\begin{table}
    \renewcommand{\tabcolsep}{2.0pt}
        \caption{Integrated line flux densities of two kinematic components and the            overall lines in the source.}
        \label{tab:lines}
        \centering
        \begin{tabular}{llllc}
        \hline
        \hline  line(CO)        &   $I_b$ (Jy\ km\ s$^{-1}$)     &       $I_r$ (Jy\ km\ s$^{-1}$)           &    I (Jy\ km\ s$^{-1}$)        &  Refs  \\  
        \hline  (1-0)     &   -                 &  -                &$0.17\pm0.06$          &  1        \\               
            (3-2)     &$1.21\pm 0.09$       &$1.22\pm0.08$      &$2.42\pm0.12$  &  -        \\
            (4-3)     &$1.29\pm 0.10$       &$1.30\pm0.09$      &$2.59\pm0.13$  &  2        \\
            (5-4)     &$1.38\pm 0.07$       &$1.41\pm0.07$      &$2.79\pm0.10$  &  4        \\
            (6-5)     &$1.34\pm 0.07$       &$1.89\pm0.07$      &$3.23\pm0.10$  &  4        \\
            (7-6)     &$1.28\pm 0.08$       &$1.51\pm0.08$      &$2.79\pm0.12$  &  4        \\
            (8-7)     &   -                 &  -                &$2.08\pm0.21$  &  1        \\ 
            (9-8)     &$0.88\pm 0.06$       &$0.95\pm0.06$      &$1.84\pm0.09$  &  3        \\
            (11-10)   &   -                 &  -                &$1.17\pm0.18$  &  1        \\ 
            
        \hline
        \end{tabular}
        \tablerefs {(1): \cite{Harrington2021}; (2): \cite{Geach2018}; (3): \cite{Riechers2021}; (4): This work}
\end{table}

\begin{table}[htbp]
    \caption{Molecular gas properties of HERS1 and the two kinematic components inferred from MCMC sampling of the Two-components LVG model}
    \label{tab:LVGresults}
    \centering
    \begin{tabular}{lccc}
    \hline
    \hline  Source  &   log($n_{\rm H_2}$)        &   log($T_{\rm k}$) &    log($N_{\rm CO}/dv$) \\
                    &   (cm$^{-3}$)              &   (K)             &    (cm$^{-2}$ km$^{-1}$ s)  \\
    \hline  HERS1 L  &   $3.75^{+0.42}_{-0.72}$   &   $1.25^{+0.25}_{-0.11}$     &            $16.09^{+0.94}_{-0.47}$ \\
            HERS1 H  &   $3.06^{+0.40}_{-0.37}$   &   $2.68^{+0.19}_{-0.23}$     &            $17.39^{+0.22}_{-0.34}$ \\
            Blue L   &   $3.01^{+0.64}_{-0.97}$   &   $1.39^{+0.19}_{-0.17}$     &            $17.14^{+0.66}_{-0.71}$ \\
            Blue H   &   $3.49^{+0.40}_{-0.50}$   &   $2.62^{+0.23}_{-0.26}$     &            $17.11^{+0.39}_{-0.38}$ \\
            Red L    &   $3.82^{+1.33}_{-1.41}$   &   $1.19^{+0.17}_{-0.11}$     &            $17.25^{+0.90}_{-1.26}$ \\
            Red H    &   $3.15^{+0.38}_{-0.32}$   &   $2.57^{+0.21}_{-0.26}$     &            $17.41^{+0.18}_{-0.26}$ \\
    \hline
    \end{tabular}
    \tablecomments{Values for each parameter are the median value and 1$\sigma$ range from the marginal probability distribution. The `L' and `H' indicate the low and high excitation. The `HERS1', `Blue' and `Red' denote the total and the two kinematic components, respectively.}
\end{table}

\subsection{FIR spectral energy distribution and inferred parameters}
We fit the SED using the publicly available SED-fitting package {\sc magphys} \citep{Cunha2008}. {\sc magphys} provides a library of model temples based on a largely empirical but physically motivated model. The stellar light is computed using the \cite{Bruzual2003} synthesis code, and the attenuation is described by a two-component model of \cite{Charlot2000} which gives the infrared luminosity absorbed and reradiated by the dust \citep{Cunha2008}. Depending on the redshift of the background, the high-$z$ extension version \citep{Cunha2015}, which extends the SED parameter priors to the high redshift, was used. 

\begin{figure}
\centering
\includegraphics[trim=2.5cm 0cm 0cm 0cm, scale=0.6]{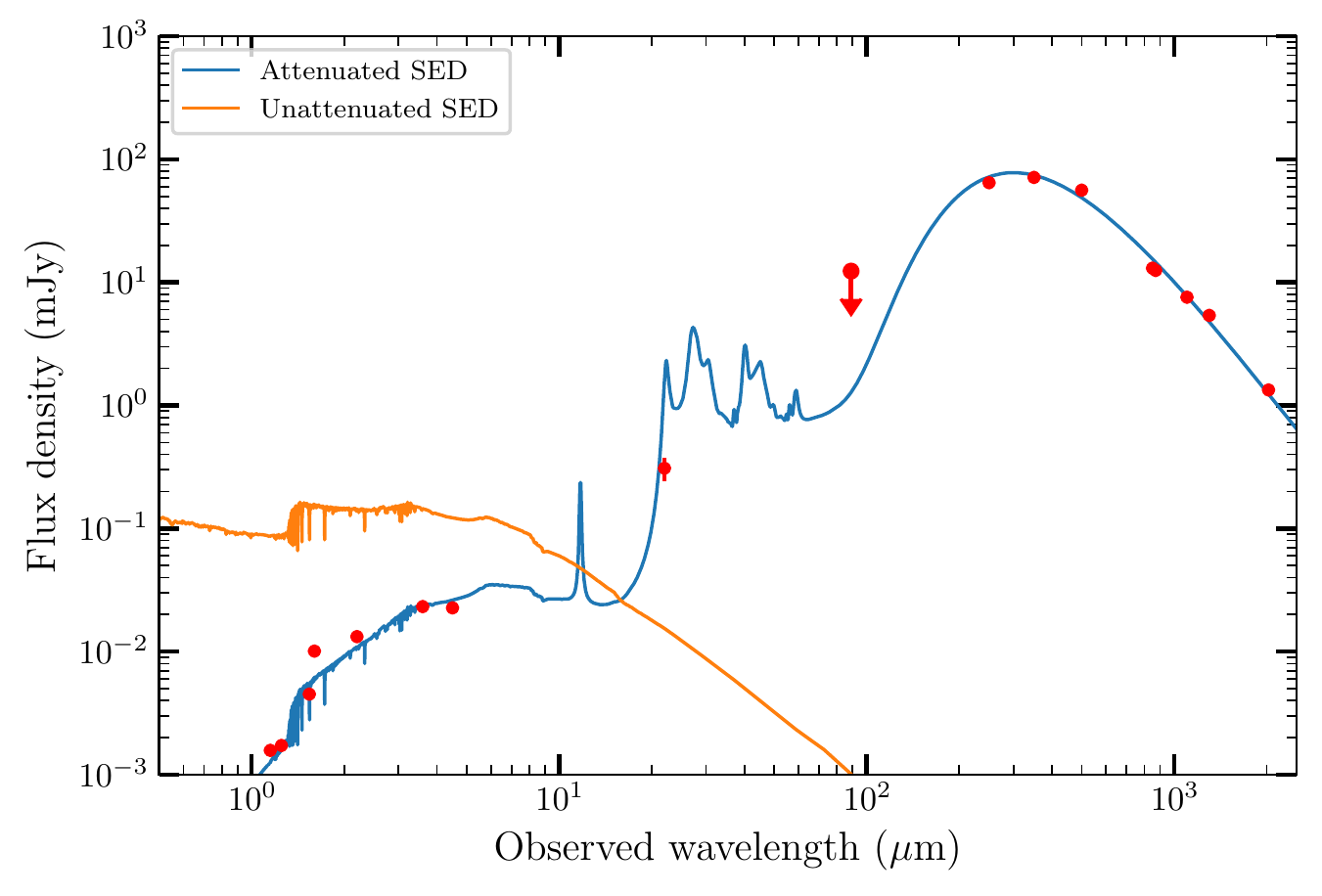}
\caption{Best-fit SED of HERS1 using the demagnified flux densities in {\sc magphys}. The blue line shows the attenuated SED while the orange line shows the unattenuated SED. The red points are the photometric results, as tabulated in Table~\ref{tab:photometry} after demagnification.}
\label{fig:SED}
\end{figure}

Multiband fluxes were required by {\sc magphys}. We used the data includeing HST/WFC3 F110W, F125W, F160W data; Keck H and $K_s$ data, Spizer/IRAC $3.6  \ \mu$m, and $4.5\ \mu $m data; the observation from Herschel/SPIRE $250\ \mu$m, $350\ \mu $m, and $500\ \mu $m; and data from SCUBA $850\ \mu $m, SMA $870\ \mu $m, AzTEC 1.1 mm, IRAM 1.3 mm, and ACT22 2.026 mm. These results are listed in Table \ref{tab:photometry}. In addition, we also include the 89 $\mu$m flux density from an observation with SOFIA/HWAC+ as an upper limit though it is not included in the model fitting to extract HERS1 physical properties. We corrected the data for lensing magnification and fixed the redshift for the source in {\sc magphys} when doing model fits to the flux density values. 
The best-fit SED result is shown in Figure \ref{fig:SED} and Table \ref{tab:sedresult}. 

The SED-fitting procedure used here did not include a contribution from an AGN. As mentioned above, \cite{Geach2015} has shown that HERS1 contains a radio-loud AGN with eMERLIN observations, showing a radio flux density excess at 1.4GHz above that expected from the far-IR-radio relation for star-forming galaxies. The presence of an AGN through CO SLED, especially the high-$J$ component, as discussed above, is inconclusive. Therefore, we expect that ignoring this AGN component does not strongly affect the final results presented here. One reason for this argument is that the 22 $\mu$m WISE/W4 band does not show a prominent excess in  the mid-IR band. Employing an AGN diagnostic suggested in \cite{Stanley2018}, the flux density ratio from the data is $log\ 10(F_{870\mu m}/F_{24\mu m}) \sim 1.5$ inferred from the SMA and WISE results. Such a value indicates HERS1 resides in the pure star-forming galaxies area and any AGN contribution to the IR luminosity is likely less than 20$\%$. For such an AGN contribution \cite{Hayward2015} has shown that {\sc magphys} can give a robust inference of galaxy properties, even if the AGN contribution to the total IR luminosity reaches up to 25$\%$. These results suggest that the lack of an AGN contribution in the SED fitting is reliable. The differential lensing effect was also considered to be negligible because all of the current continuum models have similar magnification factors.

\begin{figure}
\centering
\includegraphics[trim=2.5cm2 0cm 0cm 0cm, scale=0.4]{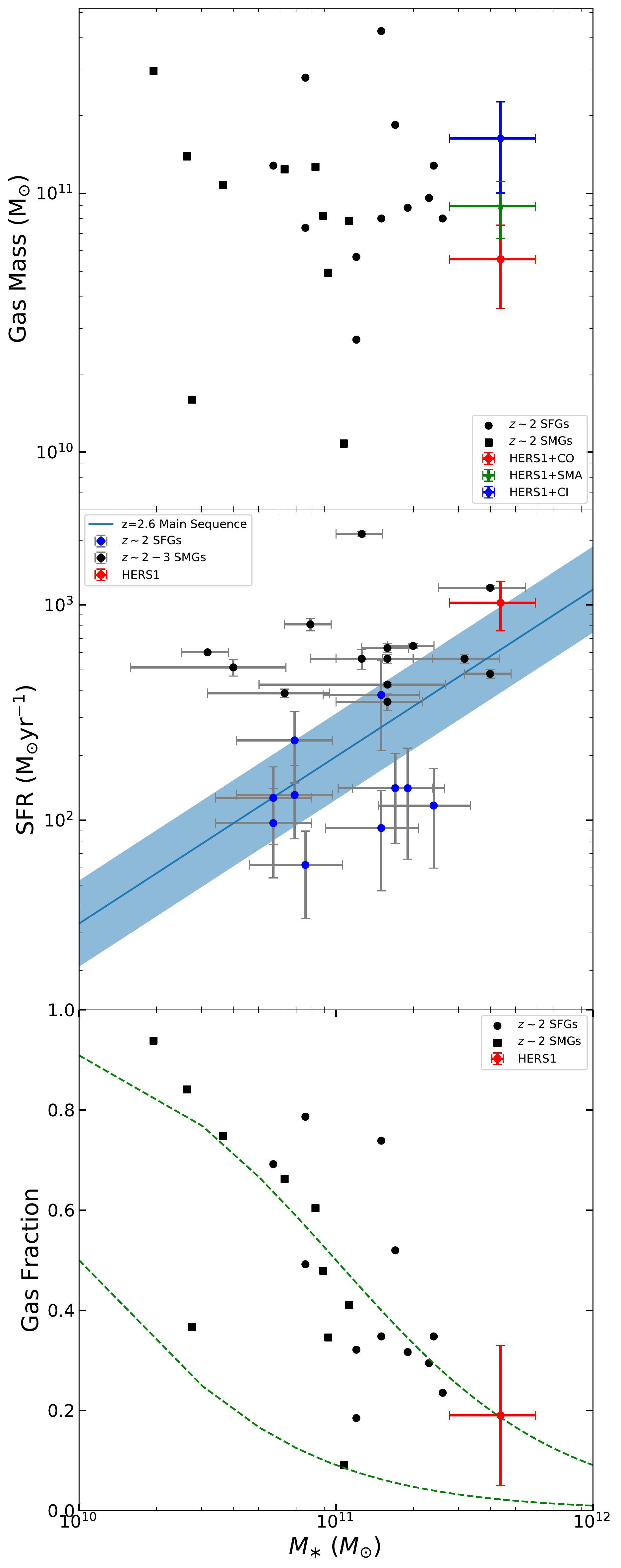}
\caption{Top: molecular gas mass calculated from CO(1-0) (red), CI lines (blue) and SMA 870\ $\mu$m (green) along with SFGs and SMGs at similar redshift \citep{Genzel2010,Bothwell2013}. Middle: stellar mass vs. star formation rate of HERS1. The blue line and shadow blue area show the $z=2.6$ trend and a 0.2 dex scatter suggested by \cite{Speagle2014}. SFGs at $z\sim2$ and SMGs at at $z\sim2-3$ \citep{Genzel2010, ivison2011,Magnelli2012} are presented for comparison. Bottom: similar with top panel but shows molecular gas fraction. The fraction of HERS1 takes the average value of three HERS1 data. The green dash lines represent a constant gas mass of $10^{10}\ {\rm M}_{\odot}$ and $10^{11}\ {\rm M}_{\odot}$, respectively. }
\label{fig:FigSFR}
\end{figure}

\begin{figure}
\centering
\includegraphics[trim=2.5cm 0cm 2.5cm 0cm, scale=0.6]{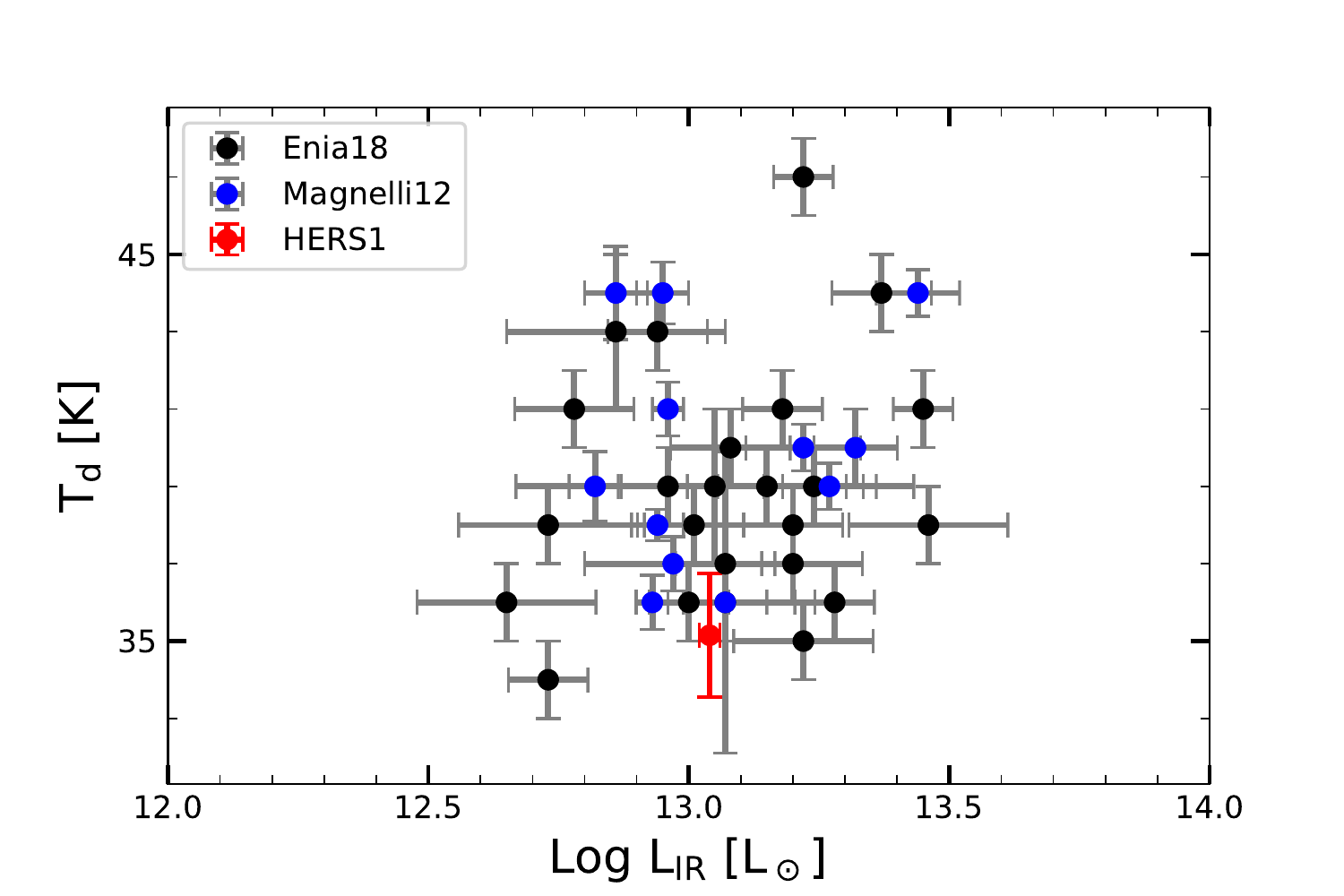}
\caption{Dust temperature vs. FIR luminosity. The $z\sim2-3$ SMGs \citep{Magnelli2012,Enia2018} are plotted for comparison .}
\label{fig:LTD}
\end{figure}

From the best-fit SED model, we can derive the total intrinsic infrared luminosity of HERS1,  $L_{\rm IR} = (1.0\pm0.3) \times 10^{13}\ L_{\odot}$, which makes it one of the hyperluminous infrared galaxies at high-redshift. The corresponding star formation rate is $1023 \pm264\ M_{\odot}$ yr$^{-1}$ assuming a Chabrier initial mass function \citep{Chabrier2003} and a convertion formula ${\rm SFR} = 1 \times 10^{-10} L_{\rm IR}$.  HERS1 also possesses a large value of stellar mass, which is in agreement with simulations \citep{Dave2010} and model requirements of submillimeter-bright galaxies \citep{Hayward2011}. The relation between stellar mass and SFR relation is shown in the top panel of Figure \ref{fig:FigSFR}. It is suggested that there is a tight correlation (called `main sequence') between the SFR and stellar mass for the majority of star-forming galaxies both in the local and high redshifts. Figure \ref{fig:FigSFR} presents the $z=2.6$ `main sequence' of \cite{Speagle2014} with a scatter of 0.2 dex. As we can see, HERS1 has a  higher SFR than the `main sequence' value at the corresponding stellar mass. The huge molecular gas reservoirs are available to meet the intense star formation activities.  For the dust temperature, the best-fit result is $T{\rm_d}=35.1^{+1.9}_{-1.4}$ K. The dust temperature is correlated with the infrared-luminosity for both local infrared luminous SMGs and high-redshift far-infrared or submillimeter-selected galaxies as shown by many works \citep{Chapman2005,Hwang2010,Magdis2010,Elbaz2011,Casey2012,Magnelli2012,Symeonidis2013,Magnelli2014,Bthermin2015}. It is a useful way to study different galaxy populations using the $L_{\rm IR}-T_{\rm d}$ relation. The dust temperature of galaxies at high redshift is likely biased to a cooler value compared with local galaxies with the same luminosities. \cite{Hwang2010} found a modest evolution of $L_{\rm IR}-T_{\rm d}$ relation as a function of redshift using {\sc Herschel}-selected samples out to $z\sim 2-3$. They concluded that SMGs are on average $2-5$ K cooler than the local counterpart from their observation. A larger scatter of the relation at high redshift is also possible, as found by \cite{Magnelli2012}, although this can be reconciled with the results of \cite{Hwang2010} by considering selection effects. Figure \ref{fig:LTD} shows the dust temperature and infrared luminosities of HERS1 with submilimeter- and Herschel-selected SMGs at $z\sim 2-3$. Compared with these galaxies, HERS1 has a colder temperature similar to other lensed candidate samples of \cite{Nayyeri2016}.

\begin{table}[htbp]
    \centering
    \caption{Photometry of HERS1}
    \label{tab:photometry}
    \centering
    \begin{tabular}{lc}
    \hline
    \hline Instrument       &     Flux Density \\
    \hline HST/WFC3 F110W      &        $21.4\pm0.8\ \mu $Jy\\
           HST/WFC3 F125W       &        $23.5\pm1.4\ \mu $Jy\\ 
           HST/WFC3 F160W       &        $61.4\pm2.2\ \mu $Jy\\
           KECK H               &        $137.3\pm1.9\ \mu $Jy\\
           KECK $K_S$           &        $180.1\pm2.3\ \mu $Jy\\
           Spizer/IRAC $3.6\ \mu$m         &  $315.8\pm12.0\ \mu $Jy\\
           Spizer/IRAC $4.5\ \mu$m         &  $308.4\pm11.9\ \mu $Jy\\
           WISE/W4                         &      4.2$\pm$ 0.9 mJy \\ 
           SOFIA/HAWC+/Band C               &    $<$ 168 mJy (3$\sigma$)\\
           Herschel/SPIRE $250\ \mu$m      &  $826\pm7 $ mJy\\
           Herschel/SPIRE $350\ \mu$m      &  $912\pm7 $ mJy\\
           Herschel/SPIRE $500\ \mu$m      &  $718\pm8 $ mJy\\
           SCUBA $850\ \mu $m                     &  $167\pm4 $ mJy\\
           SMA   $870\ \mu $m                       &  $160\pm3 $ mJy\\
           AZTEC $1.1$ mm                        &  $95.5\pm2.4 $ mJy\\
           IRAM $1.3$ mm                         &  $69\pm2.7 $ mJy\\
           ACT22 $2.026$ mm                      &  $17.1\pm1.6 $ mJy \\
    \hline
    \end{tabular}
    \tablecomments{The observed flux densities were demagnified in the SED fitting.  The HST, Keck, and Spizer data were demagnified by $13.6\pm0.4$ and other  data were demagnified by $12.8\pm0.3$.}
\end{table}

\begin{table}[htbp]
    \caption{Derived properties of HERS1 from SED fitting}
    \label{tab:sedresult}
    \centering
    \begin{tabular}{lcc}
    \hline
    \hline Quantity  &   Value  &   Unit  \\  
    \hline        $M_*$    &   $4.3^{+2.2}_{-1.0}\times 10^{11}$   &    $M_{\odot}$  \\
            $M_{\rm d}$    &   $2.6 \pm 0.3\times 10^9$               &    $M_{\odot}$  \\
            $T_{\rm d}$    &   $35.1^{+1.9}_{-1.4}$                &    K          \\
            $L_{\rm IR}$ &   $1.0\pm 0.3\times 10^{13}$            &    $L_{\odot}$  \\
            $\rm SFR$    &   $1023\pm 264$                            &    $M_{\odot}$ yr$^{-1}$ \\
    z                   & 2.553             &  ...  \\
    \hline
    \end{tabular}
\end{table}

\subsection{Molecular gas and CI}

The intrinsic luminosity $L^{'}_{\rm CO(1-0)}$ of CO(1-0) is $(5.12\pm 1.81) \times 10^{10}$ K km\ s$^{-1}$ pc$^2$ using the GBT data observed by \cite{Harrington2021} as mentioned above. The molecular gas mass is $(5.57\pm1.97) \times 10^{10}\ M_{\odot}$, adopting a conversion factor $\alpha_{\rm CO} = 0.8\ M_{\odot}$ (K\ km s$^{-1}$ pc$^2$)$^{-1}$, which is usually used in the high-redshift starburst galaxies and taking into the a factor of 1.36 for helium. We also calculated the gas mass from SMA data using the empirical calibration of \cite{Scoville2014}. The result is $(8.92\pm2.23) \times 10^{10}\ M_{\odot}$. Note that the conversion factor $\alpha_{\rm CO}$ is one of the main uncertainty sources in the molecular gas mass determination. In the CO-H$_2$ conversion, different values are taken for various types of galaxies. The factor $\alpha_{\rm CO}$=0.8 is commonly used for starburst galaxies while a high value $\alpha_{\rm CO}=4.0$ is often used in Milky Way and other local star-forming galaxies. \cite{Scoville2014} also used a high conversion factor during the calibration of $\alpha_{850}$ in their empirical relation of gas mass and dust emission. The lower $\alpha_{\rm CO}$ value adopted in this paper results a very low gas/dust ratio ($\sim 20$), which makes it become an extreme case. More research is required to determine the $\alpha_{\rm CO}$ in HERS1. For example, \cite{Genzel2015} has shown that $\alpha_{\rm CO}$ could be measured given the metallicity of the galaxy. This could help to further constrain the $\alpha_{\rm CO}$ value. The observation of CO(7-6) also had a detection of C\,I(2-1) line at 227.79 GHz `for free.' We traced the CI line into the source plane using the same method for high-$J$ CO lines. The result is shown in Figure \ref{fig:Fig highco} as well. It is suggested that C\,I lines arise from the same region as low-$J$ CO transitions and can be used to derive the gas properties without other information. The intrinsic luminosity of C\,I(2-1) from our observation is $L^{'}_{\rm CI(2-1)} = (1.60 \pm 0.22)\times 10^{10}$ K\ km\ s$^{-1}$ pc$^2$. Combing the C\,I(1-0) observation giving the intrinsic luminosity $L^{'}_{\rm CI(1-0)} = (1.65 \pm 0.20)\times 10^{10}$ K\ km\ s$^{-1}$\ pc$^2$ , we can derive the carbon excitation temperature $T_{\rm ex}=50$ K from the formula $T_{\rm ex}=38.8\ {\rm K}/{\rm ln}[2.11/R_{\rm CI}]$ where $R_{\rm CI}$ is the ratio between C\,I(2-1) and C\,I(1-0) luminosity. The carbon mass was estimated following \cite{Weib2003}, $M_{\rm CI} = (2.16\pm0.26)\times 10^7\ M_{\odot}$. The atomic carbon can also be an effective tracer to measure the gas mass. Assuming the atom carbon abundance $X[{\rm CI}]/X[{\rm H_2}] = M({\rm CI})/(6M({\rm H_2})) = 3 \times 10^{-5}$, the gas mass is $(1.63\pm0.63) \times 10^{11}\ M_{\odot}$ including helium. Figure \ref{fig:FigSFR} shows the gas mass results of the above three methods along with other galaxies at $z\sim 2$.  The carbon abundance also shows variation in different galaxies. Though there are no significant changes, a higher \cite[e.g.,][] {Weiss2005,Walter2011} or lower \cite[e.g.][]{Jiao2021} abundance will slightly alter the consistency of the results derived from CI and other components.

Having the gas mass and star formation rate, we can derive the gas depletion timescale $t_{\rm dep} \equiv M_{\rm gas}/{\rm SFR}$. HERS1 has a gas depletion of $\sim$100 Myr; this is much smaller than $\sim$1 Gyr for star-forming galaxies \cite[e.g.,][]{Kennicutt1998,Genzel2010,Saintonge2011,Decarli2016a,Decarli2016b} and a `main sequence' $\sim$0.7 Gyr or even shorter \citep{Saintonge2013,Tacconi2013,Sargent2014}. The gas fraction $f_{\rm gas}$ can be calculated as $M_{\rm gas}/(M_* + M_{\rm gas})$. HERS1 has a low gas fraction with $f_{\rm gas}= 0.19\pm0.14$ as shown in the bottom panel of Figure \ref{fig:FigSFR}. The low gas fraction and high stellar mass indicate HERS1 has formed most of its stars.

\section{Summary}
\label{section:summary}
We present a detailed study of an extremely luminous SMG at $z =2.553$, gravitationally lensed by two foreground galaxies at $z=0.202$. The lensed galaxy, dubbed HERS1, features a partial Einstein ring with a radius of $\sim$ 3$^{\prime\prime}$ observed in high-resolution maps of HST/WFC3, SMA, and ALMA. Based on the reconstructed lens model, we find magnification factors $\mu_{\rm star} = 13.6\pm 0.4$ and $\mu_{\rm dust} = 12.8 \pm 0.3$ for stellar and dust emissions of HERS1, respectively. We perform SED fitting on multiband photometry of HERS1, corrected for magnification, to measure its physical properties including stellar mass $M_*=4.3\times10^{11}\ M_{\odot}$, star formation rate SFR=$1023\ M_{\odot}$ yr$^{-1}$, and dust temperature $T_{\rm d}=35$ K. 

We analyze the physical conditions of the molecular gas through CO SLED modeling and find that its low-excitation component has a gas density $n_{\rm H_2}=10^{3.8}$ cm$^{-3}$, and kinetic temperature $T_{\rm k}=18$ K, while the high-excitation component has $n_{H_2}=10^{3.1}$cm$^{-3}$, and $T_{\rm k}=479$ K. We also find that HERS1 shows higher excitation compared to an average SMG. We further derive the total molecular gas of HERS1 using three distinct tracers, including CO(1-0), CI lines, and SMA 870\ $\mu$m. We measure a gas fraction $f_{\rm gas}=0.19$ with a depletion time $t_{\rm dep}\sim 100$ Myr. The short gas depletion time, compared to 1 Gyr for typical SFGs at $z\sim 2$, suggests that HERS1 will become quiescent shortly owing to the lack of cool gas replenishment. The location of HERS1 on the SFR-$M_*$ relation shows that it is located on the massive end of the main sequence of star formation. It reveals that HERS1 has formed the bulk of its stellar mass by $z\sim2.5$, and is about to enter a quiescent phase through the halting of cool gas replenishment, possibly caused by feedback or environmental processes.          

Moreover, we report the detection of another lensing arc feature in deep HST/WFC3 images. The feature is also detected in [N\,II] 205 $\mu $m, implying that it is at the same redshift as HERS1. We compare [N\,II] 205 $\mu $m line profile of this feature with that of HERS1 and find that the extra lensing feature has a narrower line width. We thus conclude that this extra feature is originated from a different region. Due to the lack of high S/N multiband detections, further deep observations are needed to fully understand the physical properties of this extra lensing component.

\section*{Acknowledgement} 

Financial support for this work was provided by NSF
through AST-1313319, PID 05-0087 from SOFIA for H.N. and A.C.. A.C., H.N. and N.C. acknowledge support from NASA ADAP 80NSSC20K04337 for archival data analysis of bright Herschel sources. U.C.I. group also acknowledges NASA
support for Herschel/SPIRE GTO and Open-Time Programs. This work is also based on observations made with the NASA/DLR Stratospheric Observatory for Infrared Astronomy (SOFIA). SOFIA is jointly operated by the Universities Space Research Association, Inc. (USRA), under NASA contract NAS2-97001, and the Deutsches SOFIA Institut (DSI) under DLR contract 50 OK 0901 to the University of Stuttgart. B.L. is supported by the China Scholarship Council grant (CSC No. 201906040079). Z.-H.Z. is supported by the National Natural Science Foundation of China under grants Nos. 11633001, 11920101003, and 12021003, the Strategic Priority Research Program of the Chinese Academy of Sciences, grant No. XDB23000000 and the Interdiscipline Research Funds of Beijing Normal University. E.B. and E.M.C. are supported by MIUR grant PRIN 2017 20173ML3WW-001 and Padua University grants DOR1885254/18, DOR1935272/19, and DOR2013080/20. C.Y. acknowledges support from an ESO Fellowship. This paper makes use of the following ALMA data: ADS/JAO.ALMA\#2016.2.00105.S and ADS/JAO.ALMA\#2018.1.00922.S. ALMA is a partnership of ESO (representing its member states), NSF (USA) and NINS (Japan), together with NRC (Canada), MOST and ASIAA (Taiwan), and KASI (Republic of Korea), in cooperation with the Republic of Chile. The Joint ALMA Observatory is operated by ESO, AUI/NRAO and NAOJ. The National Radio Astronomy Observatory is a facility of the National Science Foundation operated under cooperative agreement by Associated Universities, Inc. The authors wish to extend special thanks to those of Hawaiian ancestry on whose sacred mountain we are privileged to be guests. Without their generous hospitality, most of the observations presented herein would not have been possible. The Submillimeter Array is a joint project between the Smithsonian Astrophysical Observatory and the Academia Sinica Institute of Astronomy and Astrophysics and is funded by the Smithsonian Institution and the Academia Sinica.

\bibliographystyle{apj}
\bibliography{references}

\begin{thebibliography}{}
\expandafter\ifx\csname natexlab\endcsname\relax\def\natexlab#1{#1}\fi

\bibitem[{{B{\'e}thermin} {et~al.}(2015){B{\'e}thermin}, {Daddi}, {Magdis},
  {Lagos}, {Sargent}, {Albrecht}, {Aussel}, {Bertoldi}, {Buat}, {Galametz},
  {Heinis}, {Ilbert}, {Karim}, {Koekemoer}, {Lacey}, {Le Floc'h}, {Navarrete},
  {Pannella}, {Schreiber}, {Smol{\v{c}}i{\'c}}, {Symeonidis}, \&
  {Viero}}]{Bthermin2015}
{B{\'e}thermin}, M., {Daddi}, E., {Magdis}, G., {et~al.} 2015, \aap, 573, A113

\bibitem[{{Bigiel} {et~al.}(2008){Bigiel}, {Leroy}, {Walter}, {Brinks}, {de
  Blok}, {Madore}, \& {Thornley}}]{Bigiel2008}
{Bigiel}, F., {Leroy}, A., {Walter}, F., {et~al.} 2008, \aj, 136, 2846

\bibitem[{Bothwell {et~al.}(2013)Bothwell, Smail, Chapman, Genzel, Ivison,
  Tacconi, Alaghband-Zadeh, Bertoldi, Blain, Casey, Cox, Greve, Lutz, Neri,
  Omont, \& Swinbank}]{Bothwell2013}
Bothwell, M.~S., Smail, I., Chapman, S.~C., {et~al.} 2013, Monthly Notices of
  the Royal Astronomical Society, 429, 3047

\bibitem[{Bradley {et~al.}(2020)Bradley, Sipőcz, Robitaille, Tollerud,
  Vinícius, Deil, Barbary, Wilson, Busko, Günther, Cara, Conseil, Bostroem,
  Droettboom, Bray, Bratholm, Lim, Barentsen, Craig, Pascual, Perren, Greco,
  Donath, de~Val-Borro, Kerzendorf, Bach, Weaver, D'Eugenio, Souchereau, \&
  Ferreira}]{Bradley2020}
Bradley, L., Sipőcz, B., Robitaille, T., {et~al.} 2020, astropy/photutils:
  1.0.0, doi:10.5281/zenodo.4044744

\bibitem[{Bruzual \& Charlot(2003)}]{Bruzual2003}
Bruzual, G., \& Charlot, S. 2003, Monthly Notices of the Royal Astronomical
  Society, 344, 1000

\bibitem[{{Bussmann} {et~al.}(2013){Bussmann}, {P{\'e}rez-Fournon}, {Amber},
  {Calanog}, {Gurwell}, {Dannerbauer}, {De Bernardis}, {Fu}, {Harris}, {Krips},
  {Lapi}, {Maiolino}, {Omont}, {Riechers}, {Wardlow}, {Baker}, {Birkinshaw},
  {Bock}, {Bourne}, {Clements}, {Cooray}, {De Zotti}, {Dunne}, {Dye}, {Eales},
  {Farrah}, {Gavazzi}, {Gonz{\'a}lez Nuevo}, {Hopwood}, {Ibar}, {Ivison},
  {Laporte}, {Maddox}, {Mart{\'\i}nez-Navajas}, {Michalowski}, {Negrello},
  {Oliver}, {Roseboom}, {Scott}, investigate{Serjeant}, {Smith}, {Smith},
  {Streblyanska}, {Valiante}, {van der Werf}, {Verma}, {Vieira}, {Wang}, \&
  {Wilner}}]{Bussmann2013}
{Bussmann}, R.~S., {P{\'e}rez-Fournon}, I., {Amber}, S., {et~al.} 2013, \apj,
  779, 25

\bibitem[{{Ca{\~n}ameras} {et~al.}(2018){Ca{\~n}ameras}, {Yang}, {Nesvadba},
  {Beelen}, {Kneissl}, {Koenig}, {Le Floc'h}, {Limousin}, {Malhotra}, {Omont},
  \& {Scott}}]{Canameras2018}
{Ca{\~n}ameras}, R., {Yang}, C., {Nesvadba}, N.~P.~H., {et~al.} 2018, \aap,
  620, A61

\bibitem[{Carilli \& Walter(2013)}]{carilli2013}
Carilli, C., \& Walter, F. 2013, Annual Review of Astronomy and Astrophysics,
  51, 105

\bibitem[{{Carniani} {et~al.}(2019){Carniani}, {Gallerani}, {Vallini},
  {Pallottini}, {Tazzari}, {Ferrara}, {Maiolino}, {Cicone}, {Feruglio}, {Neri},
  {D'Odorico}, {Wang}, \& {Li}}]{Carniani2019}
{Carniani}, S., {Gallerani}, S., {Vallini}, L., {et~al.} 2019, \mnras, 489,
  3939

\bibitem[{{Casey}(2016)}]{Casey2016}
{Casey}, C.~M. 2016, \apj, 824, 36

\bibitem[{{Casey} {et~al.}(2014){Casey}, {Narayanan}, \& {Cooray}}]{Casey2014}
{Casey}, C.~M., {Narayanan}, D., \& {Cooray}, A. 2014, \physrep, 541, 45

\bibitem[{{Casey} {et~al.}(2012){Casey}, {Berta}, {B{\'e}thermin}, {Bock},
  {Bridge}, {Budynkiewicz}, {Burgarella}, {Chapin}, {Chapman}, {Clements},
  {Conley}, {Conselice}, {Cooray}, {Farrah}, {Hatziminaoglou}, {Ivison}, {le
  Floc'h}, {Lutz}, {Magdis}, {Magnelli}, {Oliver}, {Page}, {Pozzi},
  {Rigopoulou}, {Riguccini}, {Roseboom}, {Sanders}, {Scott}, {Seymour},
  {Valtchanov}, {Vieira}, {Viero}, \& {Wardlow}}]{Casey2012}
{Casey}, C.~M., {Berta}, S., {B{\'e}thermin}, M., {et~al.} 2012, \apj, 761, 140

\bibitem[{{Chabrier}(2003)}]{Chabrier2003}
{Chabrier}, G. 2003, \pasp, 115, 763

\bibitem[{{Chapman} {et~al.}(2005){Chapman}, {Blain}, {Smail}, \&
  {Ivison}}]{Chapman2005}
{Chapman}, S.~C., {Blain}, A.~W., {Smail}, I., \& {Ivison}, R.~J. 2005, \apj,
  622, 772

\bibitem[{Charlot \& Fall(2000)}]{Charlot2000}
Charlot, S., \& Fall, S.~M. 2000, The Astrophysical Journal, 539, 718

\bibitem[{{Cicone} {et~al.}(2014){Cicone}, {Maiolino}, {Sturm},
  {Graci{\'a}-Carpio}, {Feruglio}, {Neri}, {Aalto}, {Davies}, {Fiore},
  {Fischer}, {Garc{\'{\i}}a-Burillo}, {Gonz{\'a}lez-Alfonso},
  {Hailey-Dunsheath}, {Piconcelli}, \& {Veilleux}}]{Cicone2014}
{Cicone}, C., {Maiolino}, R., {Sturm}, E., {et~al.} 2014, \aap, 562, A21

\bibitem[{{Coil} {et~al.}(2015){Coil}, {Aird}, {Reddy}, {Shapley}, {Kriek},
  {Siana}, {Mobasher}, {Freeman}, {Price}, \& {Shivaei}}]{Coil2015}
{Coil}, A.~L., {Aird}, J., {Reddy}, N., {et~al.} 2015, \apj, 801, 35

\bibitem[{{da Cunha} {et~al.}(2008){da Cunha}, {Charlot}, \&
  {Elbaz}}]{Cunha2008}
{da Cunha}, E., {Charlot}, S., \& {Elbaz}, D. 2008, \mnras, 388, 1595

\bibitem[{{da Cunha} {et~al.}(2015){da Cunha}, {Walter}, {Smail}, {Swinbank},
  {Simpson}, {Decarli}, {Hodge}, {Weiss}, {van der Werf}, {Bertoldi},
  {Chapman}, {Cox}, {Danielson}, {Dannerbauer}, {Greve}, {Ivison}, {Karim}, \&
  {Thomson}}]{Cunha2015}
{da Cunha}, E., {Walter}, F., {Smail}, I.~R., {et~al.} 2015, \apj, 806, 110

\bibitem[{{Daddi} {et~al.}(2015){Daddi}, {Dannerbauer}, {Liu}, {Aravena},
  {Bournaud}, {Walter}, {Riechers}, {Magdis}, {Sargent}, {B{\'e}thermin},
  {Carilli}, {Cibinel}, {Dickinson}, {Elbaz}, {Gao}, {Gobat}, {Hodge}, \&
  {Krips}}]{Daddi2015}
{Daddi}, E., {Dannerbauer}, H., {Liu}, D., {et~al.} 2015, \aap, 577, A46

\bibitem[{Davé {et~al.}(2010)Davé, Finlator, Oppenheimer, Fardal, Katz,
  Kereš, \& Weinberg}]{Dave2010}
Davé, R., Finlator, K., Oppenheimer, B.~D., {et~al.} 2010, Monthly Notices of
  the Royal Astronomical Society, 404, 1355

\bibitem[{Decarli {et~al.}(2016{\natexlab{a}})Decarli, Walter, Aravena,
  Carilli, Bouwens, da~Cunha, Daddi, Ivison, Popping, Riechers, Smail,
  Swinbank, Weiss, Anguita, Assef, Bauer, Bell, Bertoldi, Chapman, Colina,
  Cortes, Cox, Dickinson, Elbaz, G{\'{o}}nzalez-L{\'{o}}pez, Ibar, Infante,
  Hodge, Karim, Fevre, Magnelli, Neri, Oesch, Ota, Rix, Sargent, Sheth, van~der
  Wel, van~der Werf, \& Wagg}]{Decarli2016a}
Decarli, R., Walter, F., Aravena, M., {et~al.} 2016{\natexlab{a}}, The
  Astrophysical Journal, 833, 69

\bibitem[{Decarli {et~al.}(2016{\natexlab{b}})Decarli, Walter, Aravena,
  Carilli, Bouwens, da~Cunha, Daddi, Elbaz, Riechers, Smail, Swinbank, Weiss,
  Bacon, Bauer, Bell, Bertoldi, Chapman, Colina, Cortes, Cox,
  G{\'{o}}nzalez-L{\'{o}}pez, Inami, Ivison, Hodge, Karim, Magnelli, Ota,
  Popping, Rix, Sargent, van~der Wel, \& van~der Werf}]{Decarli2016b}
---. 2016{\natexlab{b}}, The Astrophysical Journal, 833, 70

\bibitem[{{Doherty} {et~al.}(2020){Doherty}, {Geach}, {Ivison}, \&
  {Dye}}]{Doherty2020}
{Doherty}, M.~J., {Geach}, J.~E., {Ivison}, R.~J., \& {Dye}, S. 2020, \apj,
  905, 152

\bibitem[{{Dowell} {et~al.}(2013){Dowell}, {Staguhn}, {Harper}, {Ames},
  {Benford}, {Berthoud}, {Chapman}, {Chuss}, {Dotson}, {Irwin}, {Jhabvala},
  {Kovacs}, {Looney}, {Novak}, {Stacey}, {Vaillancourt}, \& {HAWC+ Science
  Collaboration}}]{Dowell2013}
{Dowell}, C.~D., {Staguhn}, J., {Harper}, D.~A., {et~al.} 2013, in American
  Astronomical Society Meeting Abstracts, Vol. 221, American Astronomical
  Society Meeting Abstracts \#221, 345.14

\bibitem[{{Elbaz} {et~al.}(2011){Elbaz}, {Dickinson}, {Hwang},
  {D{\'\i}az-Santos}, {Magdis}, {Magnelli}, {Le Borgne}, {Galliano},
  {Pannella}, {Chanial}, {Armus}, {Charmandaris}, {Daddi}, {Aussel}, {Popesso},
  {Kartaltepe}, {Altieri}, {Valtchanov}, {Coia}, {Dannerbauer}, {Dasyra},
  {Leiton}, {Mazzarella}, {Alexander}, {Buat}, {Burgarella}, {Chary}, {Gilli},
  {Ivison}, {Juneau}, {Le Floc'h}, {Lutz}, {Morrison}, {Mullaney}, {Murphy},
  {Pope}, {Scott}, {Brodwin}, {Calzetti}, {Cesarsky}, {Charlot}, {Dole},
  {Eisenhardt}, {Ferguson}, {F{\"o}rster Schreiber}, {Frayer}, {Giavalisco},
  {Huynh}, {Koekemoer}, {Papovich}, {Reddy}, {Surace}, {Teplitz}, {Yun}, \&
  {Wilson}}]{Elbaz2011}
{Elbaz}, D., {Dickinson}, M., {Hwang}, H.~S., {et~al.} 2011, \aap, 533, A119

\bibitem[{{Enia} {et~al.}(2018){Enia}, {Negrello}, {Gurwell}, {Dye},
  {Rodighiero}, {Massardi}, {De Zotti}, {Franceschini}, {Cooray}, {van der
  Werf}, {Birkinshaw}, {Micha{\l}owski}, \& {Oteo}}]{Enia2018}
{Enia}, A., {Negrello}, M., {Gurwell}, M., {et~al.} 2018, \mnras, 475, 3467

\bibitem[{Fixsen {et~al.}(1999)Fixsen, Bennett, \& Mather}]{Fixsen1999}
Fixsen, D.~J., Bennett, C.~L., \& Mather, J.~C. 1999, The Astrophysical
  Journal, 526, 207

\bibitem[{{F{\"o}rster Schreiber} {et~al.}(2014){F{\"o}rster Schreiber},
  {Genzel}, {Newman}, {Kurk}, {Lutz}, {Tacconi}, {Wuyts}, {Bandara}, {Burkert},
  {Buschkamp}, {Carollo}, {Cresci}, {Daddi}, {Davies}, {Eisenhauer}, {Hicks},
  {Lang}, {Lilly}, {Mainieri}, {Mancini}, {Naab}, {Peng}, {Renzini}, {Rosario},
  {Shapiro Griffin}, {Shapley}, {Sternberg}, {Tacchella}, {Vergani},
  {Wisnioski}, {Wuyts}, \& {Zamorani}}]{Forster2014}
{F{\"o}rster Schreiber}, N.~M., {Genzel}, R., {Newman}, S.~F., {et~al.} 2014,
  \apj, 787, 38

\bibitem[{{Fu} {et~al.}(2012){Fu}, {Jullo}, {Cooray}, {Bussmann}, {Ivison},
  {P{\'e}rez-Fournon}, {Djorgovski}, {Scoville}, {Yan}, {Riechers}, {Aguirre},
  {Auld}, {Baes}, {Baker}, {Bradford}, {Cava}, {Clements}, {Dannerbauer},
  {Dariush}, {De Zotti}, {Dole}, {Dunne}, {Dye}, {Eales}, {Frayer}, {Gavazzi},
  {Gurwell}, {Harris}, {Herranz}, {Hopwood}, {Hoyos}, {Ibar}, {Jarvis}, {Kim},
  {Leeuw}, {Lupu}, {Maddox}, {Mart{\'{\i}}nez-Navajas}, {Micha{\l}owski},
  {Negrello}, {Omont}, {Rosenman}, {Scott}, {Serjeant}, {Smail}, {Swinbank},
  {Valiante}, {Verma}, {Vieira}, {Wardlow}, \& {van der Werf}}]{Fu2012}
{Fu}, H., {Jullo}, E., {Cooray}, A., {et~al.} 2012, \apj, 753, 134

\bibitem[{{Gallerani} {et~al.}(2014){Gallerani}, {Ferrara}, {Neri}, \&
  {Maiolino}}]{Gallerani2014}
{Gallerani}, S., {Ferrara}, A., {Neri}, R., \& {Maiolino}, R. 2014, \mnras,
  445, 2848

\bibitem[{Geach {et~al.}(2018)Geach, Ivison, Dye, \& Oteo}]{Geach2018}
Geach, J.~E., Ivison, R.~J., Dye, S., \& Oteo, I. 2018, The Astrophysical
  Journal, 866, L12

\bibitem[{{Geach} {et~al.}(2015){Geach}, {More}, {Verma}, {Marshall},
  {Jackson}, {Belles}, {Beswick}, {Baeten}, {Chavez}, {Cornen}, {Cox}, {Erben},
  {Erickson}, {Garrington}, {Harrison}, {Harrington}, {Hughes}, {Ivison},
  {Jordan}, {Lin}, {Leauthaud}, {Lintott}, {Lynn}, {Kapadia}, {Kneib},
  {Macmillan}, {Makler}, {Miller}, {Monta{\~n}a}, {Mujica}, {Muxlow},
  {Narayanan}, {O'Briain}, {O'Brien}, {Oguri}, {Paget}, {Parrish}, {Ross},
  {Rozo}, {Rusu}, {Rykoff}, {Sanchez-Arg{\"u}elles}, {Simpson}, {Snyder},
  {Schloerb}, {Tecza}, {Wang}, {Van Waerbeke}, {Wilcox}, {Viero}, {Wilson},
  {Yun}, \& {Zeballos}}]{Geach2015}
{Geach}, J.~E., {More}, A., {Verma}, A., {et~al.} 2015, \mnras, 452, 502

\bibitem[{Genzel {et~al.}(2010)Genzel, Tacconi, Gracia-Carpio, Sternberg,
  Cooper, Shapiro, Bolatto, Bouché, Bournaud, Burkert, Combes, Comerford, Cox,
  Davis, Schreiber, Garcia-Burillo, Lutz, Naab, Neri, Omont, Shapley, \&
  Weiner}]{Genzel2010}
Genzel, R., Tacconi, L.~J., Gracia-Carpio, J., {et~al.} 2010, Monthly Notices
  of the Royal Astronomical Society, 407, 2091

\bibitem[{{Genzel} {et~al.}(2011){Genzel}, {Newman}, {Jones}, {F{\"o}rster
  Schreiber}, {Shapiro}, {Genel}, {Lilly}, {Renzini}, {Tacconi}, {Bouch{\'e}},
  {Burkert}, {Cresci}, {Buschkamp}, {Carollo}, {Ceverino}, {Davies}, {Dekel},
  {Eisenhauer}, {Hicks}, {Kurk}, {Lutz}, {Mancini}, {Naab}, {Peng},
  {Sternberg}, {Vergani}, \& {Zamorani}}]{Genzel2011}
{Genzel}, R., {Newman}, S., {Jones}, T., {et~al.} 2011, \apj, 733, 101

\bibitem[{{Genzel} {et~al.}(2015){Genzel}, {Tacconi}, {Lutz}, {Saintonge},
  {Berta}, {Magnelli}, {Combes}, {Garc{\'\i}a-Burillo}, {Neri}, {Bolatto},
  {Contini}, {Lilly}, {Boissier}, {Boone}, {Bouch{\'e}}, {Bournaud}, {Burkert},
  {Carollo}, {Colina}, {Cooper}, {Cox}, {Feruglio}, {F{\"o}rster Schreiber},
  {Freundlich}, {Gracia-Carpio}, {Juneau}, {Kovac}, {Lippa}, {Naab}, {Salome},
  {Renzini}, {Sternberg}, {Walter}, {Weiner}, {Weiss}, \& {Wuyts}}]{Genzel2015}
{Genzel}, R., {Tacconi}, L.~J., {Lutz}, D., {et~al.} 2015, \apj, 800, 20

\bibitem[{{Goldsmith}(2001)}]{Goldsmith2001}
{Goldsmith}, P.~F. 2001, \apj, 557, 736

\bibitem[{{Greve} {et~al.}(2014){Greve}, {Leonidaki}, {Xilouris}, {Wei{\ss}},
  {Zhang}, {van der Werf}, {Aalto}, {Armus}, {D{\'\i}az-Santos}, {Evans},
  {Fischer}, {Gao}, {Gonz{\'a}lez-Alfonso}, {Harris}, {Henkel}, {Meijerink},
  {Naylor}, {Smith}, {Spaans}, {Stacey}, {Veilleux}, \& {Walter}}]{Greve2014}
{Greve}, T.~R., {Leonidaki}, I., {Xilouris}, E.~M., {et~al.} 2014, \apj, 794,
  142

\bibitem[{{Hailey-Dunsheath} {et~al.}(2008){Hailey-Dunsheath}, {Nikola},
  {Stacey}, {Oberst}, {Parshley}, {Bradford}, {Ade}, \& {Tucker}}]{Hailey2008}
{Hailey-Dunsheath}, S., {Nikola}, T., {Stacey}, G.~J., {et~al.} 2008, \apjl,
  689, L109

\bibitem[{{Harrington} {et~al.}(2016){Harrington}, {Yun}, {Cybulski}, {Wilson},
  {Aretxaga}, {Chavez}, {De la Luz}, {Erickson}, {Ferrusca}, {Gallup},
  {Hughes}, {Monta{\~n}a}, {Narayanan}, {S{\'a}nchez-Arg{\"u}elles},
  {Schloerb}, {Souccar}, {Terlevich}, {Terlevich}, {Zeballos}, \&
  {Zavala}}]{Harrington2016}
{Harrington}, K.~C., {Yun}, M.~S., {Cybulski}, R., {et~al.} 2016, \mnras, 458,
  4383

\bibitem[{Harrington {et~al.}(2019)Harrington, Vishwas, Weiß, Magnelli,
  Grassitelli, Zajaček, Jiménez-Andrade, Leung, Bertoldi, Romano-Díaz,
  Frayer, Kamieneski, Riechers, Stacey, Yun, \& Wang}]{Harrington2019}
Harrington, K.~C., Vishwas, A., Weiß, A., {et~al.} 2019, Monthly Notices of
  the Royal Astronomical Society, 488, 1489

\bibitem[{{Harrington} {et~al.}(2021){Harrington}, {Weiss}, {Yun}, {Magnelli},
  {Sharon}, {Leung}, {Vishwas}, {Wang}, {Frayer}, {Jim{\'e}nez-Andrade}, {Liu},
  {Garc{\'\i}a}, {Romano-D{\'\i}az}, {Frye}, {Jarugula}, {B{\u{a}}descu},
  {Berman}, {Dannerbauer}, {D{\'\i}az-S{\'a}nchez}, {Grassitelli},
  {Kamieneski}, {Kim}, {Kirkpatrick}, {Lowenthal}, {Messias}, {Puschnig},
  {Stacey}, {Torne}, \& {Bertoldi}}]{Harrington2021}
{Harrington}, K.~C., {Weiss}, A., {Yun}, M.~S., {et~al.} 2021, \apj, 908, 95

\bibitem[{Hayward {et~al.}(2011)Hayward, Kere{\v{s}}, Jonsson, Narayanan, Cox,
  \& Hernquist}]{Hayward2011}
Hayward, C.~C., Kere{\v{s}}, D., Jonsson, P., {et~al.} 2011, The Astrophysical
  Journal, 743, 159

\bibitem[{{Hayward} \& {Smith}(2015)}]{Hayward2015}
{Hayward}, C.~C., \& {Smith}, D. J.~B. 2015, \mnras, 446, 1512

\bibitem[{{Herrera-Camus} {et~al.}(2016){Herrera-Camus}, {Bolatto}, {Smith},
  {Draine}, {Pellegrini}, {Wolfire}, {Croxall}, {de Looze}, {Calzetti},
  {Kennicutt}, {Crocker}, {Armus}, {van der Werf}, {Sandstrom}, {Galametz},
  {Brandl}, {Groves}, {Rigopoulou}, {Walter}, {Leroy}, {Boquien}, {Tabatabaei},
  \& {Beirao}}]{Herrera2016}
{Herrera-Camus}, R., {Bolatto}, A., {Smith}, J.~D., {et~al.} 2016, \apj, 826,
  175

\bibitem[{{Ho} {et~al.}(2004){Ho}, {Moran}, \& {Lo}}]{Ho2004}
{Ho}, P. T.~P., {Moran}, J.~M., \& {Lo}, K.~Y. 2004, \apjl, 616, L1

\bibitem[{{Hwang} {et~al.}(2010){Hwang}, {Elbaz}, {Magdis}, {Daddi},
  {Symeonidis}, {Altieri}, {Amblard}, {Andreani}, {Arumugam}, {Auld}, {Aussel},
  {Babbedge}, {Berta}, {Blain}, {Bock}, {Bongiovanni}, {Boselli}, {Buat},
  {Burgarella}, {Castro-Rodr{\'\i}guez}, {Cava}, {Cepa}, {Chanial}, {Chapin},
  {Chary}, {Cimatti}, {Clements}, {Conley}, {Conversi}, {Cooray},
  {Dannerbauer}, {Dickinson}, {Dominguez}, {Dowell}, {Dunlop}, {Dwek}, {Eales},
  {Farrah}, {F{\"o}rster Schreiber}, {Fox}, {Franceschini}, {Gear}, {Genzel},
  {Glenn}, {Griffin}, {Gruppioni}, {Halpern}, {Hatziminaoglou}, {Ibar},
  {Isaak}, {Ivison}, {Jeong}, {Lagache}, {Le Borgne}, {Le Floc'h}, {Lee},
  {Lee}, {Lee}, {Levenson}, {Lu}, {Lutz}, {Madden}, {Maffei}, {Magnelli},
  {Mainetti}, {Maiolino}, {Marchetti}, {Mortier}, {Nguyen}, {Nordon},
  {O'Halloran}, {Okumura}, {Oliver}, {Omont}, {Page}, {Panuzzo},
  {Papageorgiou}, {Pearson}, {P{\'e}rez-Fournon}, {Garc{\'\i}a}, {Poglitsch},
  {Pohlen}, {Popesso}, {Pozzi}, {Rawlings}, {Rigopoulou}, {Riguccini}, {Rizzo},
  {Rodighiero}, {Roseboom}, {Rowan-Robinson}, {Saintonge}, {Portal}, {Santini},
  {Sauvage}, {Schulz}, {Scott}, {Seymour}, {Shao}, {Shupe}, {Smith}, {Stevens},
  {Sturm}, {Tacconi}, {Trichas}, {Tugwell}, {Vaccari}, {Valtchanov}, {Vieira},
  {Vigroux}, {Wang}, {Ward}, {Wright}, {Xu}, \& {Zemcov}}]{Hwang2010}
{Hwang}, H.~S., {Elbaz}, D., {Magdis}, G., {et~al.} 2010, \mnras, 409, 75

\bibitem[{{Ivison} {et~al.}(2011{\natexlab{a}}){Ivison}, {Papadopoulos},
  {Smail}, {Greve}, {Thomson}, {Xilouris}, \& {Chapman}}]{ivison2011}
{Ivison}, R.~J., {Papadopoulos}, P.~P., {Smail}, I., {et~al.}
  2011{\natexlab{a}}, \mnras, 412, 1913

\bibitem[{{Ivison} {et~al.}(2011{\natexlab{b}}){Ivison}, {Papadopoulos},
  {Smail}, {Greve}, {Thomson}, {Xilouris}, \& {Chapman}}]{Ivison2011b}
---. 2011{\natexlab{b}}, Monthly Notices of the Royal Astronomical Society,
  412, 1913

\bibitem[{{Jarugula} {et~al.}(2021){Jarugula}, {Vieira}, {Weiss}, {Spilker},
  {Aravena}, {Archipley}, {B{\'e}thermin}, {Chapman}, {Dong}, {Greve},
  {Harrington}, {Hayward}, {Hezaveh}, {Hill}, {Litke}, {Malkan}, {Marrone},
  {Narayanan}, {Phadke}, {Reuter}, \& {Rotermund}}]{Jarugula2021}
{Jarugula}, S., {Vieira}, J.~D., {Weiss}, A., {et~al.} 2021, \apj, 921, 97

\bibitem[{{Jiao} {et~al.}(2021){Jiao}, {Gao}, \& {Zhao}}]{Jiao2021}
{Jiao}, Q., {Gao}, Y., \& {Zhao}, Y. 2021, \mnras, 504, 2360

\bibitem[{{Kennicutt}(1998)}]{Kennicutt1998}
{Kennicutt}, Robert~C., J. 1998, \apj, 498, 541

\bibitem[{{Kov{\'a}cs}(2008)}]{Kovacs2008}
{Kov{\'a}cs}, A. 2008, in Society of Photo-Optical Instrumentation Engineers
  (SPIE) Conference Series, Vol. 7020, Millimeter and Submillimeter Detectors
  and Instrumentation for Astronomy IV, ed. W.~D. {Duncan}, W.~S. {Holland},
  S.~{Withington}, \& J.~{Zmuidzinas}, 70201S

\bibitem[{{Kriek} {et~al.}(2015){Kriek}, {Shapley}, {Reddy}, {Siana}, {Coil},
  {Mobasher}, {Freeman}, {de Groot}, {Price}, {Sanders}, {Shivaei}, {Brammer},
  {Momcheva}, {Skelton}, {van Dokkum}, {Whitaker}, {Aird}, {Azadi}, {Kassis},
  {Bullock}, {Conroy}, {Dav{\'e}}, {Kere{\v s}}, \& {Krumholz}}]{Kriek2015}
{Kriek}, M., {Shapley}, A.~E., {Reddy}, N.~A., {et~al.} 2015, \apjs, 218, 15

\bibitem[{{Lu} {et~al.}(2014){Lu}, {Zhao}, {Xu}, {Gao}, {Armus}, {Mazzarella},
  {Isaak}, {Petric}, {Charmandaris}, {D{\'\i}az-Santos}, {Evans}, {Howell},
  {Appleton}, {Inami}, {Iwasawa}, {Leech}, {Lord}, {Sanders}, {Schulz},
  {Surace}, \& {van der Werf}}]{Lu2014}
{Lu}, N., {Zhao}, Y., {Xu}, C.~K., {et~al.} 2014, \apjl, 787, L23

\bibitem[{{Lu} {et~al.}(2017){Lu}, {Zhao}, {D{\'\i}az-Santos}, {Xu}, {Gao},
  {Armus}, {Isaak}, {Mazzarella}, {van der Werf}, {Appleton}, {Charmandaris},
  {Evans}, {Howell}, {Iwasawa}, {Leech}, {Lord}, {Petric}, {Privon}, {Sanders},
  {Schulz}, \& {Surace}}]{Lu2017}
{Lu}, N., {Zhao}, Y., {D{\'\i}az-Santos}, T., {et~al.} 2017, \apjs, 230, 1

\bibitem[{{Ma} {et~al.}(2018){Ma}, {Brown}, {Cooray}, {Nayyeri}, {Messias},
  {Timmons}, {Staguhn}, {Temi}, {Dowell}, {Wardlow}, {Fadda}, {Kovacs},
  {Riechers}, {Oteo}, {Wilson}, \& {Perez-Fournon}}]{Ma2018}
{Ma}, J., {Brown}, A., {Cooray}, A., {et~al.} 2018, \apj, 864, 60

\bibitem[{{Magdis} {et~al.}(2010){Magdis}, {Elbaz}, {Hwang}, {Amblard},
  {Arumugam}, {Aussel}, {Blain}, {Bock}, {Boselli}, {Buat},
  {Castro-Rodr{\'\i}guez}, {Cava}, {Chanial}, {Clements}, {Conley}, {Conversi},
  {Cooray}, {Dowell}, {Dwek}, {Eales}, {Farrah}, {Franceschini}, {Glenn},
  {Griffin}, {Halpern}, {Hatziminaoglou}, {Huang}, {Ibar}, {Isaak}, {Le
  Floc'h}, {Lagache}, {Levenson}, {Lonsdale}, {Lu}, {Madden}, {Maffei},
  {Mainetti}, {Marchetti}, {Morrison}, {Nguyen}, {O'Halloran}, {Oliver},
  {Omont}, {Owen}, {Page}, {Pannella}, {Panuzzo}, {Papageorgiou}, {Pearson},
  {P{\'e}rez-Fournon}, {Pohlen}, {Rigopoulou}, {Rizzo}, {Roseboom},
  {Rowan-Robinson}, {Schulz}, {Scott}, {Seymour}, {Shupe}, {Smith}, {Stevens},
  {Strazzullo}, {Symeonidis}, {Trichas}, {Tugwell}, {Vaccari}, {Valtchanov},
  {Vigroux}, {Wang}, {Wright}, {Xu}, \& {Zemcov}}]{Magdis2010}
{Magdis}, G.~E., {Elbaz}, D., {Hwang}, H.~S., {et~al.} 2010, \mnras, 409, 22

\bibitem[{{Magnelli} {et~al.}(2012){Magnelli}, {Lutz}, {Santini}, {Saintonge},
  {Berta}, {Albrecht}, {Altieri}, {Andreani}, {Aussel}, {Bertoldi},
  {B{\'e}thermin}, {Bongiovanni}, {Capak}, {Chapman}, {Cepa}, {Cimatti},
  {Cooray}, {Daddi}, {Danielson}, {Dannerbauer}, {Dunlop}, {Elbaz}, {Farrah},
  {F{\"o}rster Schreiber}, {Genzel}, {Hwang}, {Ibar}, {Ivison}, {Le Floc'h},
  {Magdis}, {Maiolino}, {Nordon}, {Oliver}, {P{\'e}rez Garc{\'\i}a},
  {Poglitsch}, {Popesso}, {Pozzi}, {Riguccini}, {Rodighiero}, {Rosario},
  {Roseboom}, {Salvato}, {Sanchez-Portal}, {Scott}, {Smail}, {Sturm},
  {Swinbank}, {Tacconi}, {Valtchanov}, {Wang}, \& {Wuyts}}]{Magnelli2012}
{Magnelli}, B., {Lutz}, D., {Santini}, P., {et~al.} 2012, \aap, 539, A155

\bibitem[{{Magnelli} {et~al.}(2014){Magnelli}, {Lutz}, {Saintonge}, {Berta},
  {Santini}, {Symeonidis}, {Altieri}, {Andreani}, {Aussel}, {B{\'e}thermin},
  {Bock}, {Bongiovanni}, {Cepa}, {Cimatti}, {Conley}, {Daddi}, {Elbaz},
  {F{\"o}rster Schreiber}, {Genzel}, {Ivison}, {Le Floc'h}, {Magdis},
  {Maiolino}, {Nordon}, {Oliver}, {Page}, {P{\'e}rez Garc{\'\i}a}, {Poglitsch},
  {Popesso}, {Pozzi}, {Riguccini}, {Rodighiero}, {Rosario}, {Roseboom},
  {Sanchez-Portal}, {Scott}, {Sturm}, {Tacconi}, {Valtchanov}, {Wang}, \&
  {Wuyts}}]{Magnelli2014}
{Magnelli}, B., {Lutz}, D., {Saintonge}, A., {et~al.} 2014, \aap, 561, A86

\bibitem[{Mashian {et~al.}(2015)Mashian, Sturm, Sternberg, Janssen,
  Hailey-Dunsheath, Fischer, Contursi, Gonz{\'{a}}lez-Alfonso,
  Graci{\'{a}}-Carpio, Poglitsch, Veilleux, Davies, Genzel, Lutz, Tacconi,
  Verma, Wei{\ss}, Polisensky, \& Nikola}]{Mashian2015}
Mashian, N., Sturm, E., Sternberg, A., {et~al.} 2015, The Astrophysical
  Journal, 802, 81

\bibitem[{{M{\'e}ndez-Abreu} {et~al.}(2008){M{\'e}ndez-Abreu}, {Aguerri},
  {Corsini}, \& {Simonneau}}]{Mendez2008}
{M{\'e}ndez-Abreu}, J., {Aguerri}, J.~A.~L., {Corsini}, E.~M., \& {Simonneau},
  E. 2008, \aap, 478, 353

\bibitem[{{M{\'e}ndez-Abreu} {et~al.}(2017){M{\'e}ndez-Abreu}, {Ruiz-Lara},
  {S{\'a}nchez-Menguiano}, {de Lorenzo-C{\'a}ceres}, {Costantin},
  {Catal{\'a}n-Torrecilla}, {Florido}, {Aguerri}, {Bland-Hawthorn}, {Corsini},
  {Dettmar}, {Galbany}, {Garc{\'\i}a-Benito}, {Marino}, {M{\'a}rquez},
  {Ortega-Minakata}, {Papaderos}, {S{\'a}nchez}, {S{\'a}nchez-Blazquez},
  {Spekkens}, {van de Ven}, {Wild}, \& {Ziegler}}]{Mendez2017}
{M{\'e}ndez-Abreu}, J., {Ruiz-Lara}, T., {S{\'a}nchez-Menguiano}, L., {et~al.}
  2017, \aap, 598, A32

\bibitem[{{Men{\'e}ndez-Delmestre} {et~al.}(2013){Men{\'e}ndez-Delmestre},
  {Blain}, {Swinbank}, {Smail}, {Ivison}, {Chapman}, \& {Gon{\c
  c}alves}}]{Menendez2013}
{Men{\'e}ndez-Delmestre}, K., {Blain}, A.~W., {Swinbank}, M., {et~al.} 2013,
  \apj, 767, 151

\bibitem[{Nayyeri {et~al.}(2016)Nayyeri, Keele, Cooray, Riechers, Ivison,
  Harris, Frayer, Baker, Chapman, Eales, Farrah, Fu, Marchetti, Marques-Chaves,
  Martinez-Navajas, Oliver, Omont, Perez-Fournon, Scott, Vaccari, Vieira,
  Viero, Wang, \& Wardlow}]{Nayyeri2016}
Nayyeri, H., Keele, M., Cooray, A., {et~al.} 2016, The Astrophysical Journal,
  823, 17

\bibitem[{{Negrello} {et~al.}(2007){Negrello}, {Perrotta},
  {Gonz{\'a}lez-Nuevo}, {Silva}, {de Zotti}, {Granato}, {Baccigalupi}, \&
  {Danese}}]{Negrello2007}
{Negrello}, M., {Perrotta}, F., {Gonz{\'a}lez-Nuevo}, J., {et~al.} 2007,
  \mnras, 377, 1557

\bibitem[{{Negrello} {et~al.}(2010){Negrello}, {Hopwood}, {De Zotti}, {Cooray},
  {Verma}, {Bock}, {Frayer}, {Gurwell}, {Omont}, {Neri}, {Dannerbauer},
  {Leeuw}, {Barton}, {Cooke}, {Kim}, {da Cunha}, {Rodighiero}, {Cox},
  {Bonfield}, {Jarvis}, {Serjeant}, {Ivison}, {Dye}, {Aretxaga}, {Hughes},
  {Ibar}, {Bertoldi}, {Valtchanov}, {Eales}, {Dunne}, {Driver}, {Auld},
  {Buttiglione}, {Cava}, {Grady}, {Clements}, {Dariush}, {Fritz}, {Hill},
  {Hornbeck}, {Kelvin}, {Lagache}, {Lopez-Caniego}, {Gonzalez-Nuevo}, {Maddox},
  {Pascale}, {Pohlen}, {Rigby}, {Robotham}, {Simpson}, {Smith}, {Temi},
  {Thompson}, {Woodgate}, {York}, {Aguirre}, {Beelen}, {Blain}, {Baker},
  {Birkinshaw}, {Blundell}, {Bradford}, {Burgarella}, {Danese}, {Dunlop},
  {Fleuren}, {Glenn}, {Harris}, {Kamenetzky}, {Lupu}, {Maddalena}, {Madore},
  {Maloney}, {Matsuhara}, {Micha{\l}owski}, {Murphy}, {Naylor}, {Nguyen},
  {Popescu}, {Rawlings}, {Rigopoulou}, {Scott}, {Scott}, {Seibert}, {Smail},
  {Tuffs}, {Vieira}, {van der Werf}, \& {Zmuidzinas}}]{Negrello2010}
{Negrello}, M., {Hopwood}, R., {De Zotti}, G., {et~al.} 2010, Science, 330, 800

\bibitem[{{Negrello} {et~al.}(2017){Negrello}, {Amber}, {Amvrosiadis}, {Cai},
  {Lapi}, {Gonzalez-Nuevo}, {De Zotti}, {Furlanetto}, {Maddox}, {Allen},
  {Bakx}, {Bussmann}, {Cooray}, {Covone}, {Danese}, {Dannerbauer}, {Fu},
  {Greenslade}, {Gurwell}, {Hopwood}, {Koopmans}, {Napolitano}, {Nayyeri},
  {Omont}, {Petrillo}, {Riechers}, {Serjeant}, {Tortora}, {Valiante}, {Verdoes
  Kleijn}, {Vernardos}, {Wardlow}, {Baes}, {Baker}, {Bourne}, {Clements},
  {Crawford}, {Dye}, {Dunne}, {Eales}, {Ivison}, {Marchetti}, {Micha{\l}owski},
  {Smith}, {Vaccari}, \& {van der Werf}}]{Negrello2017}
{Negrello}, M., {Amber}, S., {Amvrosiadis}, A., {et~al.} 2017, \mnras, 465,
  3558

\bibitem[{{Nikola} {et~al.}(2011){Nikola}, {Stacey}, {Brisbin}, {Ferkinhoff},
  {Hailey-Dunsheath}, {Parshley}, \& {Tucker}}]{Nikola2011}
{Nikola}, T., {Stacey}, G.~J., {Brisbin}, D., {et~al.} 2011, \apj, 742, 88

\bibitem[{Papadopoulos {et~al.}(2012)Papadopoulos, van~der Werf, Xilouris,
  Isaak, Gao, \& Mühle}]{Papadopoulos2012}
Papadopoulos, P.~P., van~der Werf, P.~P., Xilouris, E.~M., {et~al.} 2012,
  Monthly Notices of the Royal Astronomical Society, 426, 2601

\bibitem[{{Peng} {et~al.}(2010){Peng}, {Ho}, {Impey}, \& {Rix}}]{Peng2010}
{Peng}, C.~Y., {Ho}, L.~C., {Impey}, C.~D., \& {Rix}, H.-W. 2010, \aj, 139,
  2097

\bibitem[{{Rajan}(2011)}]{Rajan2011}
{Rajan}, A. 2011, in WFC3 Data Handbook v. 2.1, Vol.~2

\bibitem[{{Riechers} {et~al.}(2021){Riechers}, {Cooray}, {P{\'e}rez-Fournon},
  \& {Neri}}]{Riechers2021}
{Riechers}, D.~A., {Cooray}, A., {P{\'e}rez-Fournon}, I., \& {Neri}, R. 2021,
  \apj, 913, 141

\bibitem[{{Riechers} {et~al.}(2013){Riechers}, {Bradford}, {Clements},
  {Dowell}, {P{\'e}rez-Fournon}, {Ivison}, {Bridge}, {Conley}, {Fu}, {Vieira},
  {Wardlow}, {Calanog}, {Cooray}, {Hurley}, {Neri}, {Kamenetzky}, {Aguirre},
  {Altieri}, {Arumugam}, {Benford}, {B{\'e}thermin}, {Bock}, {Burgarella},
  {Cabrera-Lavers}, {Chapman}, {Cox}, {Dunlop}, {Earle}, {Farrah}, {Ferrero},
  {Franceschini}, {Gavazzi}, {Glenn}, {Solares}, {Gurwell}, {Halpern},
  {Hatziminaoglou}, {Hyde}, {Ibar}, {Kov{\'a}cs}, {Krips}, {Lupu}, {Maloney},
  {Martinez-Navajas}, {Matsuhara}, {Murphy}, {Naylor}, {Nguyen}, {Oliver},
  {Omont}, {Page}, {Petitpas}, {Rangwala}, {Roseboom}, {Scott}, {Smith},
  {Staguhn}, {Streblyanska}, {Thomson}, {Valtchanov}, {Viero}, {Wang},
  {Zemcov}, \& {Zmuidzinas}}]{Riechers2013}
{Riechers}, D.~A., {Bradford}, C.~M., {Clements}, D.~L., {et~al.} 2013, \nat,
  496, 329

\bibitem[{Rivera {et~al.}(2019)Rivera, Baker, Gallardo, Gralla, Harris,
  Huffenberger, Hughes, Keeton, L{\'{o}}pez-Caraballo, Marriage, Partridge,
  Sievers, Tagore, Walter, Wei{\ss}, \& Wollack}]{Rivera2019}
Rivera, J., Baker, A.~J., Gallardo, P.~A., {et~al.} 2019, The Astrophysical
  Journal, 879, 95

\bibitem[{{Rosenberg} {et~al.}(2015){Rosenberg}, {van der Werf}, {Aalto},
  {Armus}, {Charmandaris}, {D{\'\i}az-Santos}, {Evans}, {Fischer}, {Gao},
  {Gonz{\'a}lez-Alfonso}, {Greve}, {Harris}, {Henkel}, {Israel}, {Isaak},
  {Kramer}, {Meijerink}, {Naylor}, {Sanders}, {Smith}, {Spaans}, {Spinoglio},
  {Stacey}, {Veenendaal}, {Veilleux}, {Walter}, {Wei{\ss}}, {Wiedner}, {van der
  Wiel}, \& {Xilouris}}]{Rosenberg2015}
{Rosenberg}, M.~J.~F., {van der Werf}, P.~P., {Aalto}, S., {et~al.} 2015, \apj,
  801, 72

\bibitem[{Saintonge {et~al.}(2011)Saintonge, Kauffmann, Wang, Kramer, Tacconi,
  Buchbender, Catinella, Graciá-Carpio, Cortese, Fabello, Fu, Genzel,
  Giovanelli, Guo, Haynes, Heckman, Krumholz, Lemonias, Li, Moran,
  Rodriguez-Fernandez, Schiminovich, Schuster, \& Sievers}]{Saintonge2011}
Saintonge, A., Kauffmann, G., Wang, J., {et~al.} 2011, Monthly Notices of the
  Royal Astronomical Society, 415, 61

\bibitem[{{Saintonge} {et~al.}(2013){Saintonge}, {Lutz}, {Genzel}, {Magnelli},
  {Nordon}, {Tacconi}, {Baker}, {Bandara}, {Berta}, {F{\"o}rster Schreiber},
  {Poglitsch}, {Sturm}, {Wuyts}, \& {Wuyts}}]{Saintonge2013}
{Saintonge}, A., {Lutz}, D., {Genzel}, R., {et~al.} 2013, \apj, 778, 2

\bibitem[{Sargent {et~al.}(2014)Sargent, Daddi, B{\'{e}}thermin, Aussel,
  Magdis, Hwang, Juneau, Elbaz, \& da~Cunha}]{Sargent2014}
Sargent, M.~T., Daddi, E., B{\'{e}}thermin, M., {et~al.} 2014, The
  Astrophysical Journal, 793, 19

\bibitem[{{Sch{\"o}ier} {et~al.}(2005){Sch{\"o}ier}, {van der Tak}, {van
  Dishoeck}, \& {Black}}]{schoier2005}
{Sch{\"o}ier}, F.~L., {van der Tak}, F.~F.~S., {van Dishoeck}, E.~F., \&
  {Black}, J.~H. 2005, \aap, 432, 369

\bibitem[{{Scoville} {et~al.}(2014){Scoville}, {Aussel}, {Sheth}, {Scott},
  {Sanders}, {Ivison}, {Pope}, {Capak}, {Vanden Bout}, {Manohar}, {Kartaltepe},
  {Robertson}, \& {Lilly}}]{Scoville2014}
{Scoville}, N., {Aussel}, H., {Sheth}, K., {et~al.} 2014, \apj, 783, 84

\bibitem[{{Serjeant}(2012)}]{Serjeant2012}
{Serjeant}, S. 2012, \mnras, 424, 2429

\bibitem[{{Shapley} {et~al.}(2015){Shapley}, {Reddy}, {Kriek}, {Freeman},
  {Sanders}, {Siana}, {Coil}, {Mobasher}, {Shivaei}, {Price}, \& {de
  Groot}}]{Shapley2015}
{Shapley}, A.~E., {Reddy}, N.~A., {Kriek}, M., {et~al.} 2015, \apj, 801, 88

\bibitem[{{Speagle} {et~al.}(2014){Speagle}, {Steinhardt}, {Capak}, \&
  {Silverman}}]{Speagle2014}
{Speagle}, J.~S., {Steinhardt}, C.~L., {Capak}, P.~L., \& {Silverman}, J.~D.
  2014, \apjs, 214, 15

\bibitem[{Spilker {et~al.}(2014)Spilker, Marrone, Aguirre, Aravena, Ashby,
  B{\'{e}}thermin, Bradford, Bothwell, Brodwin, Carlstrom, Chapman, Crawford,
  de~Breuck, Fassnacht, Gonzalez, Greve, Gullberg, Hezaveh, Holzapfel, Husband,
  Ma, Malkan, Murphy, Reichardt, Rotermund, Stalder, Stark, Strandet, Vieira,
  Wei{\ss}, \& Welikala}]{Spilker2014}
Spilker, J.~S., Marrone, D.~P., Aguirre, J.~E., {et~al.} 2014, The
  Astrophysical Journal, 785, 149

\bibitem[{{Stanley} {et~al.}(2018){Stanley}, {Harrison}, {Alexander},
  {Simpson}, {Knudsen}, {Mullaney}, {Rosario}, \& {Scholtz}}]{Stanley2018}
{Stanley}, F., {Harrison}, C.~M., {Alexander}, D.~M., {et~al.} 2018, \mnras,
  478, 3721

\bibitem[{{Sturm} {et~al.}(2011){Sturm}, {Gonz{\'a}lez-Alfonso}, {Veilleux},
  {Fischer}, {Graci{\'a}-Carpio}, {Hailey-Dunsheath}, {Contursi}, {Poglitsch},
  {Sternberg}, {Davies}, {Genzel}, {Lutz}, {Tacconi}, {Verma}, {Maiolino}, \&
  {de Jong}}]{Sturm2011}
{Sturm}, E., {Gonz{\'a}lez-Alfonso}, E., {Veilleux}, S., {et~al.} 2011, \apjl,
  733, L16

\bibitem[{{Su} {et~al.}(2017){Su}, {Marriage}, {Asboth}, {Baker}, {Bond},
  {Crichton}, {Devlin}, {D{\"u}nner}, {Farrah}, {Frayer}, {Gralla}, {Hall},
  {Halpern}, {Harris}, {Hilton}, {Hincks}, {Hughes}, {Niemack}, {Page},
  {Partridge}, {Rivera}, {Scott}, {Sievers}, {Thornton}, {Viero}, {Wang},
  {Wollack}, \& {Zemcov}}]{su2017}
{Su}, T., {Marriage}, T.~A., {Asboth}, V., {et~al.} 2017, \mnras, 464, 968

\bibitem[{{Swinbank} {et~al.}(2011){Swinbank}, {Papadopoulos}, {Cox}, {Krips},
  {Ivison}, {Smail}, {Thomson}, {Neri}, {Richard}, \& {Ebeling}}]{Swinbank2011}
{Swinbank}, A.~M., {Papadopoulos}, P.~P., {Cox}, P., {et~al.} 2011, \apj, 742,
  11

\bibitem[{Symeonidis {et~al.}(2013)Symeonidis, Vaccari, Berta, Page, Lutz,
  Arumugam, Aussel, Bock, Boselli, Buat, Capak, Clements, Conley, Conversi,
  Cooray, Dowell, Farrah, Franceschini, Giovannoli, Glenn, Griffin,
  Hatziminaoglou, Hwang, Ibar, Ilbert, Ivison, Floc'h, Lilly, Kartaltepe,
  Magnelli, Magdis, Marchetti, Nguyen, Nordon, O’Halloran, Oliver, Omont,
  Papageorgiou, Patel, Pearson, Pérez-Fournon, Pohlen, Popesso, Pozzi,
  Rigopoulou, Riguccini, Rosario, Roseboom, Rowan-Robinson, Salvato, Schulz,
  Scott, Seymour, Shupe, Smith, Valtchanov, Wang, Xu, Zemcov, \&
  Wuyts}]{Symeonidis2013}
Symeonidis, M., Vaccari, M., Berta, S., {et~al.} 2013, Monthly Notices of the
  Royal Astronomical Society, 431, 2317

\bibitem[{{Tacconi} {et~al.}(2013){Tacconi}, {Neri}, {Genzel}, {Combes},
  {Bolatto}, {Cooper}, {Wuyts}, {Bournaud}, {Burkert}, {Comerford}, {Cox},
  {Davis}, {F{\"o}rster Schreiber}, {Garc{\'{\i}}a-Burillo}, {Gracia-Carpio},
  {Lutz}, {Naab}, {Newman}, {Omont}, {Saintonge}, {Shapiro Griffin}, {Shapley},
  {Sternberg}, \& {Weiner}}]{Tacconi2013}
{Tacconi}, L.~J., {Neri}, R., {Genzel}, R., {et~al.} 2013, \apj, 768, 74

\bibitem[{{Timmons} {et~al.}(2015){Timmons}, {Cooray}, {Nayyeri}, {Casey},
  {Calanog}, {Ma}, {Messias}, {Baes}, {Bussmann}, {Dunne}, {Dye}, {Eales},
  {Fu}, {Ivison}, {Maddox}, {Micha{\l}owski}, {Oteo}, {Riechers}, {Valiante},
  \& {Wardlow}}]{Timmons2015}
{Timmons}, N., {Cooray}, A., {Nayyeri}, H., {et~al.} 2015, \apj, 805, 140

\bibitem[{{Toft} {et~al.}(2014){Toft}, {Smol{\v c}i{\'c}}, {Magnelli}, {Karim},
  {Zirm}, {Michalowski}, {Capak}, {Sheth}, {Schawinski}, {Krogager}, {Wuyts},
  {Sanders}, {Man}, {Lutz}, {Staguhn}, {Berta}, {Mccracken}, {Krpan}, \&
  {Riechers}}]{Toft2014}
{Toft}, S., {Smol{\v c}i{\'c}}, V., {Magnelli}, B., {et~al.} 2014, \apj, 782,
  68

\bibitem[{{Tunnard} \& {Greve}(2016)}]{Tunnard2016}
{Tunnard}, R., \& {Greve}, T.~R. 2016, \apj, 819, 161

\bibitem[{{van der Tak} {et~al.}(2007){van der Tak}, {Black}, {Sch{\"o}ier},
  {Jansen}, \& {van Dishoeck}}]{vandertak2007}
{van der Tak}, F.~F.~S., {Black}, J.~H., {Sch{\"o}ier}, F.~L., {Jansen}, D.~J.,
  \& {van Dishoeck}, E.~F. 2007, \aap, 468, 627

\bibitem[{{Viero} {et~al.}(2014){Viero}, {Asboth}, {Roseboom}, {Moncelsi},
  {Marsden}, {Mentuch Cooper}, {Zemcov}, {Addison}, {Baker}, {Beelen}, {Bock},
  {Bridge}, {Conley}, {Devlin}, {Dor{\'e}}, {Farrah}, {Finkelstein},
  {Font-Ribera}, {Geach}, {Gebhardt}, {Gill}, {Glenn}, {Hajian}, {Halpern},
  {Jogee}, {Kurczynski}, {Lapi}, {Negrello}, {Oliver}, {Papovich}, {Quadri},
  {Ross}, {Scott}, {Schulz}, {Somerville}, {Spergel}, {Vieira}, {Wang}, \&
  {Wechsler}}]{Viero2014}
{Viero}, M.~P., {Asboth}, V., {Roseboom}, I.~G., {et~al.} 2014, \apjs, 210, 22

\bibitem[{{Walter} {et~al.}(2011){Walter}, {Wei{\ss}}, {Downes}, {Decarli}, \&
  {Henkel}}]{Walter2011}
{Walter}, F., {Wei{\ss}}, A., {Downes}, D., {Decarli}, R., \& {Henkel}, C.
  2011, \apj, 730, 18

\bibitem[{{Wang} {et~al.}(2019){Wang}, {Wang}, {Fan}, {Wu}, {Yang}, {Neri}, \&
  {Yue}}]{Wang2019}
{Wang}, F., {Wang}, R., {Fan}, X., {et~al.} 2019, \apj, 880, 2

\bibitem[{{Wardlow} {et~al.}(2013){Wardlow}, {Cooray}, {De Bernardis},
  {Amblard}, {Arumugam}, {Aussel}, {Baker}, {B{\'e}thermin}, {Blundell},
  {Bock}, {Boselli}, {Bridge}, {Buat}, {Burgarella}, {Bussmann},
  {Cabrera-Lavers}, {Calanog}, {Carpenter}, {Casey}, {Castro-Rodr{\'{\i}}guez},
  {Cava}, {Chanial}, {Chapin}, {Chapman}, {Clements}, {Conley}, {Cox},
  {Dowell}, {Dye}, {Eales}, {Farrah}, {Ferrero}, {Franceschini}, {Frayer},
  {Frazer}, {Fu}, {Gavazzi}, {Glenn}, {Gonz{\'a}lez Solares}, {Griffin},
  {Gurwell}, {Harris}, {Hatziminaoglou}, {Hopwood}, {Hyde}, {Ibar}, {Ivison},
  {Kim}, {Lagache}, {Levenson}, {Marchetti}, {Marsden}, {Martinez-Navajas},
  {Negrello}, {Neri}, {Nguyen}, {O'Halloran}, {Oliver}, {Omont}, {Page},
  {Panuzzo}, {Papageorgiou}, {Pearson}, {P{\'e}rez-Fournon}, {Pohlen},
  {Riechers}, {Rigopoulou}, {Roseboom}, {Rowan-Robinson}, {Schulz}, {Scott},
  {Scoville}, {Seymour}, {Shupe}, {Smith}, {Streblyanska}, {Strom},
  {Symeonidis}, {Trichas}, {Vaccari}, {Vieira}, {Viero}, {Wang}, {Xu}, {Yan},
  \& {Zemcov}}]{Wardlow2013}
{Wardlow}, J.~L., {Cooray}, A., {De Bernardis}, F., {et~al.} 2013, \apj, 762,
  59

\bibitem[{{Wei{\ss}} {et~al.}(2005){Wei{\ss}}, {Downes}, {Henkel}, \&
  {Walter}}]{Weiss2005}
{Wei{\ss}}, A., {Downes}, D., {Henkel}, C., \& {Walter}, F. 2005, \aap, 429,
  L25

\bibitem[{{Wei{\ss}} {et~al.}(2007){Wei{\ss}}, {Downes}, {Neri}, {Walter},
  {Henkel}, {Wilner}, {Wagg}, \& {Wiklind}}]{Weiss2007}
{Wei{\ss}}, A., {Downes}, D., {Neri}, R., {et~al.} 2007, \aap, 467, 955

\bibitem[{{Wei{\ss}} {et~al.}(2003){Wei{\ss}}, {Henkel}, {Downes}, \&
  {Walter}}]{Weib2003}
{Wei{\ss}}, A., {Henkel}, C., {Downes}, D., \& {Walter}, F. 2003, \aap, 409,
  L41

\bibitem[{{Yang} {et~al.}(2017){Yang}, {Omont}, {Beelen}, {Gao}, {van der
  Werf}, {Gavazzi}, {Zhang}, {Ivison}, {Lehnert}, {Liu}, {Oteo},
  {Gonz{\'a}lez-Alfonso}, {Dannerbauer}, {Cox}, {Krips}, {Neri}, {Riechers},
  {Baker}, {Micha{\l}owski}, {Cooray}, \& {Smail}}]{yang2017}
{Yang}, C., {Omont}, A., {Beelen}, A., {et~al.} 2017, \aap, 608, A144

\end{thebibliography}

\end{document}